\documentclass[a4paper,onecolumn,11pt,]{quantumarticle}
\pdfoutput=1
\usepackage[utf8]{inputenc}
\usepackage[english]{babel}
\usepackage[T1]{fontenc}
\usepackage{hyperref}

\usepackage[pdftex]{graphicx}
\usepackage{amsmath, amssymb}
\usepackage{braket}
\usepackage[numbers]{natbib}
\usepackage{color}
\usepackage{lipsum}
\usepackage{comment}
\usepackage{qcircuit}
\usepackage{pgfplots}
\usepackage{tikz} 
\usetikzlibrary{spy}
\usepgfplotslibrary{groupplots}
\usepackage{caption}
\usepackage{bm}
\usepackage{subcaption}

\newcommand{\eg}{\emph{e.g.}}

\begin{document}
\title{From Ad-Hoc to Systematic: A Strategy for Imposing General Boundary Conditions in Discretized PDEs in variational quantum algorithm}
\author{Lu Dingjie$^{\dagger *}$}
\affiliation{Institute of High Performance Computing, Agency for Science, Technology and Research (A*STAR), 138632 Singapore}

\author{Wang Zhao$^*$}
\affiliation{Institute of High Performance Computing, Agency for Science, Technology and Research (A*STAR), 138632 Singapore}

\author{Liu Jun}
\affiliation{Institute of High Performance Computing, Agency for Science, Technology and Research (A*STAR), 138632 Singapore}

\author{Li Yangfan}
\affiliation{Institute of High Performance Computing, Agency for Science, Technology and Research (A*STAR), 138632 Singapore}

\author{Wei-Bin Ewe}
\affiliation{Institute of High Performance Computing, Agency for Science, Technology and Research (A*STAR), 138632 Singapore}

\author{Liu Zhuangjian$^{\ddagger}$}
\affiliation{Institute of High Performance Computing, Agency for Science, Technology and Research (A*STAR), 138632 Singapore}

\footnotetext{$\dagger$~: ludj@ihpc.a-star.edu.sg}
\footnotetext{$\ddagger$~: liuzj@ihpc.a-star.edu.sg}
\footnotetext{*~: These authors contributed equally to this work.}

\date{03 Nov. 2023}
\begin{abstract}
We proposed a general quantum-computing-based algorithm that harnesses the exponential power of noisy intermediate-scale quantum (NISQ) devices in solving partial differential equations (PDEs). This variational quantum eigensolver (VQE)-inspired approach transcends previous idealized model demonstrations constrained by strict and simplistic boundary conditions. It enables the imposition of arbitrary boundary conditions, significantly expanding its potential and adaptability for real-world applications, achieving this ``from ad-hoc to systematic" concept. We have implemented this method using the fourth-order PDE (the Euler-Bernoulli beam) as an example and showcased its effectiveness with four different boundary conditions. This framework enables expectation evaluations independent of problem size, harnessing the exponentially growing state space inherent in quantum computing, resulting in exceptional scalability. This method paves the way for applying quantum computing to practical engineering applications.
\end{abstract}
\maketitle
\section{Introduction}\label{sec:2_intro}
Partial differential equations (PDEs) play an important role in elucidating the dynamics of numerous physical phenomena encompassing acoustics, heat transfer, fluid flow, electromagnetism, structural mechanics, and solid mechanics \citep{Renardy2006SSBM,Avalos2019MN,Rabczuk2019CMC,Brion2023CS}. Efficiently addressing these problems, especially on a large scale, holds a profound significance, since enhanced computational efficiency boosts the integration of numerical simulations and real-world experiments to advance the engineering progress. Despite some strides in settling large-scale PDE problems using classical computers \citep{Klawonn2010ZAMM,Klawonn2015SIAM,Toivanen2018IJNME}, the computational efficiency for large systems remains a formidable challenge even with the supercomputers. For instance, Fugaku, with a state-of-the-art solver capable of handling 49 billions of degrees of freedom (DOFs), still takes half a day to simulate seismic response \citep{Fujita2021JOCS}. However, this duration fails to satisfy real-time or near-real-time requirements for early warning and rapid decision-making systems. To circumvent this limitation, quantum computing has emerged as an attractive approach, offering the potential for substantial reductions in computational costs and improved parallel scalability compared to classical methods.

Quantum computing is a new computational paradigm that harnesses the principles of quantum mechanics, namely superposition and entanglement, to resolve intricate problems believed to be classically hard \citep{Horowitz2019,Haug2021PRX}. Plenty of research works have showcased the exponential speedup and favorable computational accuracy of quantum algorithms over classical counterparts. Notable examples include Shor's algorithm for integer factorization \citep{Shor1994}, solving linear systems with sparse coefficient matrices using Harrow-Hassidim-Lloyd (HHL) algorithm \citep{Harrow2009PRL} and the generalized HHL algorithm \citep{Cao2012MP,Clader2013PRL,Zhang2021IEEE}, and tackling the Poisson equation in various contexts with Quantum Phase Estimation (QPE) algorithms\citep{Cao2013NJP} or Variational quantum algorithms (VQAs) \citep{Sato2021PRA,Liu2021PRA}. While quantum algorithms will manifest supremacy once fault-tolerant quantum computers with sufficient qubits and error-correcting capabilities are available, it is believed to be a long time before such a quantum computer advents \citep{Bravo2019arXiv,Preskill2018Quantum}. At this stage, the era of noisy intermediate-scale quantum (NISQ) devices as a stepping stone towards supremacy \citep{Preskill2018Quantum} unfolds, where quantum algorithms can be executed with a limited number of qubits and a shallow quantum circuit. Unlike other algorithms, VQAs are more adaptive to NISQ devices, making them a more realistic approach for achieving tangible quantum advantages at the current juncture \citep{Cerezo2021NRP,Enomoto2022PRR}. VQAs encode a problem into a cost function in terms of the expectation values of specific observables. A parameterized quantum circuit (PQC) referred to as an ansatz, is employed by a quantum computer or simulator to evaluate this function \citep{Holmes2022PRX}. Through a series of iterations using this low-depth circuit, the results from each iteration are fed into a classical optimizer. This process continually updates the optimized parameters, thereby reducing error buildup and maximizing the capabilities of NISQ devices.

VQAs represent a hybrid quantum-classical algorithm paragon \citep{Fedorov2022MT,Cervera2021PRX,Magann2021PRX}. The Variational Quantum Eigensolver (VQE) \citep{Peruzzo2014Nature}, one of the most influential members in VQA family, encodes the system into a Hamiltonian and finds out its ground state via ansatz. Given the practical systems usually feature sparse Hamiltonians, the computational cost scales polynomially with the problem size, making it more efficient than classical methods \citep{Cerezo2021NRP,Tilly2022PR}. Another representative VQA is the Quantum Approximate Optimization Algorithm (QAOA) \citep{Farhi2014arXiv}, which is intended for combinatorial optimization problems \citep{Hadfield2019Algorithms,Alexeev2021PRX}. Apart from VQE addressing problems that pertain to continuous variables, QAOA is dedicated to identifying the optimal solution from a set of discrete possibilities, such as MaxCut, the traveling salesperson problems \citep{Hadfield2019Algorithms,Zhou2020PRX,Matos2021PRX}.

VQAs have demonstrated its capabilities on solving the linear problems \citep{Xu2021SB,Huang2021NJP,Liu2021PRA,Sato2021PRA}. For instance, Liu \emph{et al.} \cite{Liu2021PRA} explicitly decomposed a system matrix of the Poisson equation from the finite difference method (FDM) to reduce the quantum resource. This method only involves $O(\text{log} N)$ measurements, where $N$ denotes the problem size. However, this approach only produces normalized solution without the norm of the solutions. The absence of the norm leads to significant information loss in practical engineering applications \citep{Sato2021PRA}. Furthermore, Sato \emph{et al.}~\cite{Sato2021PRA} enhanced Liu \emph{et al.}'s work \cite{Liu2021PRA} by employing explicit decomposition for basic observables, hence, achieving scalability and reducing computational and measurement times. Moreover, this approach is able to calculate solution norms, thus advancing the utilization of quantum computing in computer-aided engineering. However, its widespread application is hindered by its simplistic boundary conditions (BCs). Ewe \emph{et al.}~\cite{Ewe2022IEEE} synergized VQA and FDM to solve the Helmholtz equation, successfully describing the propagation modes of electromagnetic waves in a hollow metallic waveguide. This study represents a promising step towards practical applications.

Numerical discretization strategies such as finite element method (FEM), finite difference method (FDM) or finite volume method (FVM) can complement quantum computing from several aspects, 1) it helps generate linear systems algorithmically, enabling efficient data access and potential scalability; 2) it inherently results in sparse linear systems \citep{Manteuffel1980MOC}, which is well-suited for quantum implementation and potential acceleration, as evidenced in \eg, \cite{Harrow2009PRL}; 3) it bears important and broad practical engineering relevance \citep{Rao2005} and accommodate a broader range of geometries and BCs, thus increasing the likelihood that practical problems reap the benefits of quantum speedup conjugate with these methods.

In this paper, we propose a general quantum-computing-based approach to solve PDEs, leveraging the power of NISQ devices. Apart from previous attempts constrained by well-formed and simplistic boundary conditions, our method inspired by VQE is capable of handling arbitrary boundary conditions. Our approach is validated by the fourth-order PDE (Euler-Bernoulli beam theory) with four distinct boundary conditions. The problem is numerically discretized with FEM method. The independence of expectation values on problem size enables it to exploit the exponential state space offered by quantum computing, exhibiting excellent scalability. Therefore, this framework has significant potential for real-world engineering applications, making it promising to the field.

This paper is structured as follows. Section \ref{sec:2_intro} provides a comprehensive review of the relevant literature, highlighting existing accomplishments and research gaps. In Section \ref{sec:3_method}, the methodology is delineated, detailing PDE discretization, Pauli decomposition and the imposition of BCs, followed by the corresponding implementation in Section \ref{sec:4_implementation}. The core of our study, presented in Section \ref{sec:5_results}, includes the results and empirical findings, together with a discussion and interpretation of these results. Last but not least, this paper is concluded in Section \ref{sec:6_conclusion} by summarizing our main findings, identifying limitations, and suggesting avenues for future research.
\section{Formulation of minimization problem}\label{sec:3_method}
Consider the analysis of a problem defined by PDE as~\cite{Bathe1982},
\begin{equation}
    L_{2m}[u] = r,
    \label{eq:L}
\end{equation}
in which $L_{2m}$ is a linear differential operator, $u$ is the state variable, and $r$ is the forcing term. The solution must satisfy the boundary conditions
\begin{equation}
    B_i[u] = q_i |_{S_i}; i = 1, 2, \dots,
\end{equation}
where $q_i$ is a prescribed function that the solution must satisfy at boundary
$S_i$.

Specifically, we focus on symmetric and positive definite operators that satisfy
\begin{equation}
    \int_{\Omega} {({L_{2m}}[u])v}\ d\Omega = \int_{\Omega} {({L_{2m}}[v])u}\ d\Omega\ , \quad
    \int_{\Omega} {({L_{2m}}[u])u}\ d\Omega > 0,
\end{equation}
where $\Omega$ is the domain of the operator and $u$ and $v$ are any functions that satisfy the boundary conditions.

By introducing the test function or the variation function $v$ and integrating by parts, since the operator $L_{2m}$ is symmetric, we obtain the total potential energy functional $\Pi$ as
\begin{equation}
    \Pi = \int_{\Omega} {{M_{m}}[u]}M_{m}[v]\ d\Omega - \int_{\Omega} {rv}\ d\Omega + BC,
    \label{eq:Pi}
\end{equation}
where $M_m$ is a $m$-order linear differential operator on the state variable $u$ and $BC$ is the boundary condition terms. $v \in V$ is the test function in the space defined as
\begin{equation}
    V = \{ v \in H^d(\Omega) ~|~ v = 0 ~ \mathrm{on} ~ \Gamma_\mathrm{D} \},
\end{equation}
where $d$ is the dimension of the domain $\Omega$ and $H^d(\Omega)$ is a Sobolev space~\cite{Courant2008MMP}.

According to the variational principle, the solution of the PDE described in Eq.~(\ref{eq:L}) corresponds to the stationary condition in Eq.~(\ref{eq:Pi}). Hence, Eq.~(\ref{eq:L}) can be solved by minimizing the total potential energy functional $\Pi$ on $v$.

To formulate the solution of the PDE into a minimization problem, the potential $\Pi$, which is the loss function, can be discretized with proper approaches such as FDM or FEM as
\begin{equation}
    \Pi_{d} = \frac{1}{2}{\bm{v}^T}K\bm{v} - \frac{1}{2}{\bm{v}^T}\bm{f} -\frac{1}{2}{\bm{f}^T}\bm{v},
    \label{eq:Pid}
\end{equation}
where $\Pi_d$ is the discretized potential energy, $K$ is a positive-definite symmetric matrix, often referred to as the ``stiffness matrix", $\bm{f}$ incorporates the terms $\int_{\Omega} {rv}\ d\Omega + BC$ in Eq.~(\ref{eq:Pi}) and $\bm{v}$ denotes the vector of the state variable $u$ at the nodes in the discretized domain.

To minimize $\Pi_d$ on a quantum computer with VQE, the variables need to be represented by quantum states with amplitude encoding. Since the PDE is linear, $\| f \| = 1$ is assumed, hence $\bm{f}$ can be denoted by a quantum state $\ket{f}$. In general, the vector $\bm{v}$ is not inherently a quantum state, but it can be encoded into a quantum state in a scaled manner as $\bm{v} = c\ket{\phi}$ by introducing a scaling factor $c$ and a quantum state $\ket{\phi}$, where $\ket{\phi}$ is the normalized $\bm{v}$ and $c$ is the norm. The quantum state $\ket{\phi}$ is prepared by a PQC as $\ket{\phi(\bm{\theta})}$, where $\bm{\theta}$ are the parameters.

Substituting $v = c \ket{\phi(\bm{\theta})}$ and $\bm{f} = \ket{f}$ into Eq.~(\ref{eq:Pid}) and recall that $\bm{v}$ and $\bm{f}$ are both of real amplitude/value, the total energy can be rearranged into
\begin{eqnarray}
    \Pi_d = \frac{1}{2} c^2 \bra{\phi(\bm{\theta})} K \ket{\phi(\bm{\theta})}
    - \frac{1}{2} c \braket{\phi(\bm{\theta})|{f}} - \frac{1}{2} c \braket{f | \phi(\bm{\theta})}.
\end{eqnarray}

Introducing the quantum state $\ket{f, \phi(\bm{\theta})} = \left(
\ket{0}\ket{f} + \ket{1}\ket{\phi(\bm{\theta})} \right) / \sqrt{2}$
from~\cite{Sato2021PRA, Liu2021PRA}, the total potential energy is simplified to
\begin{eqnarray}
    \Pi_d(c, \bm{\theta}) &=& \frac{1}{2} c^2 \bra{\phi(\bm{\theta})} K \ket{\phi(\bm{\theta})} - c \braket{f, \phi(\bm{\theta}) | X \otimes I^{\otimes n} | f, \phi(\bm{\theta})},
    \label{eq:PiQ}
\end{eqnarray}
where $I$ and $X$ are the matrix representation of Identity and NOT gates, respectively, and $n$ is the number of qubits where $n = \log_2 N$, with $N$ the total number of DOFs in current discretization.

Given that $c$ and $\bm{\theta}$ parameterize the vector $\bm{v}$, the process of minimizing the total potential energy with respect to $\bm{v}$ is equivalent to the minimization of the function $\Pi_d(c, \bm{\theta})$ in Eq.~(\ref{eq:PiQ}), where $c$ and $\bm{\theta}$ are the variables being optimized as
\begin{equation}
    \min_{c, \bm{\theta}} \Pi_d(c, \bm{\theta}).
    \label{eq:Obj}
\end{equation}

Letting $c^*$ and $\bm{\theta}^*$ denote the optimal solution of the minimization problem, then the vector that encodes the solution of the PDE is
$\bm{v} = c^*\ket{\phi(\bm{\theta}^*)}$.

Minimizing $\Pi_d$ requires a sequential optimization of $c$ and $\bm{\theta}$, i.e.,~\cite{Sato2021PRA},
\begin{equation}
    \min_{c, \bm{\theta}} \Pi_d(c, \bm{\theta}) = \min_{\bm{\theta}} \Pi_d(c^*(\bm{\theta}), \bm{\theta}),
\end{equation}
where $c^*(\bm{\theta})$ is the optimal value of $c$ for a given $\bm{\theta}$.

Since $\Pi_d$ is quadratic on $c$, it is straightforward to derive $c^*(\bm{\theta})$ from the vertex as,
\begin{equation}
    c^*(\bm{\theta}) = \frac{\bra{f, \phi(\bm{\theta})} X \otimes I^{\otimes n} \ket{f, \phi(\bm{\theta})}}
    {\bra{\phi(\bm{\theta})} K \ket{\phi(\bm{\theta})}}.
    \label{eq:vertex}
\end{equation}

Sequentially, the minimization of $\Pi_d$ on $\bm{\theta}$ involves substituting $c^*(\bm{\theta})$ in Eq.~(\ref{eq:vertex}) into Eq.~(\ref{eq:PiQ}), thereby eliminating $c$ from the loss function as
\begin{eqnarray}
    \Pi_d(c^*(\bm{\theta}), \bm{\theta}) =-\frac{\left( \bra{f, \phi(\bm{\theta})} X
    \otimes I^{\otimes n} \ket{ f, \phi(\bm{\theta})} \right)^2 }{2 \bra{\phi(\bm{\theta})}
    K \ket{\phi(\bm{\theta})} }.
    \label{eq:PiQt}
\end{eqnarray}
\section{Implementation}\label{sec:4_implementation}
To evaluate the loss in Eq.~(\ref{eq:PiQt}) on a quantum computer, it is crucial to break down the matrix $K$ into observables. The most direct method for achieving this involves decomposing $K$ over the Pauli basis $\mathcal P_n = \big\{ P_1 \otimes \ldots \otimes P_n : \forall i,\ P_i \in \{I, X, Y, Z\}\big\}$ as
\begin{equation}
   K = \sum_{P \in \mathcal P_n}{c_P P} ,
\end{equation}
where $c_P$ are the coefficients in the decomposition and the Pauli matrices~\cite{Nielsen2010QCQI} are defined by
\begin{align} \label{eq:pauli_gates}
    I=\ket{0}\bra{0}+\ket{1}\bra{1},\
    X=\ket{1}\bra{0}+\ket{0}\bra{1},\ Z=\ket{0}\bra{0}-\ket{1}\bra{1},\ Y=iXZ.
\end{align}

The key challenge in this approach revolves around the efficient decomposition of matrix $K$, as conventional decomposition results in number of Pauli terms that scales exponentially with the dimension of $K$, rendering this method impractical. To address this challenge, we adopt an approach akin to Sato \emph{et al.}~\cite{Sato2021PRA}. This method maximizes the utilization of the internal structure of matrix $K$, capitalizing on its repetitive block patterns. The complete matrix, formed by repeating blocks, can be expressed using tensor products. Consequently, matrix $K$ can be decomposed into a linear combination of observables within $\mathcal P_n$. The terms from this decomposition remain constant, independent of the dimension of $K$, resulting in an exponential reduction in the number of observables to be computed during the expectation calculation.

\subsection{Evaluation of $\bra{\psi} K \ket{\psi}$} \label{sec:K_exp}
To illustrate the method, we examine the equilibrium equation of Euler-Bernoulli beam, subjected to a distributed load $q$, where $w$, $E$ and $I$ represent the deflection, the Young's modulus and the second moment of inertia, respectively,
\begin{equation}
    EI\frac{{{d^4}w}}{{d{x^4}}} = q,
\end{equation}
which is discretized by FEM \cite{Bathe1982} with $N$ nodes and element size of $l_e$. In Euler-Bernoulli beam theory, each node has two DOFs, namely deflection and rotation, and each element has two nodes. For a system with $N$ DOFs, only $\log_2({N})$ qubits are required in quantum computing, which is a great space advantage compared with classical computing. The element stiffness matrix $K_e$ can be Pauli decomposed into constant (six) terms as,
\begin{eqnarray}
    K_e &=& c_1 I \otimes I + c_2 I \otimes Z + c_3 X \otimes I + c_4 X
    \otimes Z +c_5 Y \otimes Y + c_6 Z \otimes X.
    \label{eq:Ke}
\end{eqnarray}

The matrix $K$ in Eq.~(\ref{eq:PiQt}) is the composite of $K_e$ via matrix shifts. It can be partitioned into three interconnected components associated with $K_e$ as
\begin{eqnarray}\label{eq:K012}
    K = K_0 + K_1 + K_2
\end{eqnarray}
The matrices $K_e$, $K$, $K_0$, $K_1$ and $K_2$ are detailed in Appendix~\ref{app_sec:K}.

The matrix $K_0$ can be written into Pauli matrices coupled with $K_e$ as $K_0 = I^{\otimes n} \otimes K_e$. By diagonally shifting $K_0$ twice to the top-left, $K_1$ can be represented as $K_1 = P^{-1}K_0 P$, with $P$ being the shift operator~\cite{Zhang2014SCIS} as $P = \sum_{i \in [0, 2^n-1]} \ket{(i+1)\ \rm{mod}~2^n} \bra{i}$. Similarly, the $K_2$ matrix is expressed as $K_2 = P^{-1}(I_{0}^{\otimes n} \otimes K_e)P$, where $I_0$ represents a Hermitian operator defined as $I_0=\ket{0}\bra{0}$. The Pauli decomposition of $K$ ultimately comprises constant terms, regardless of the discretized problem size $N$, indicating that only $\mathcal{O}(1)$ terms are computed on a quantum computer.

Since the shift operator $P$ is unitary, the expectations of observables tied to $P$ can be evaluated directly. The initial ansatz yields $\ket{\phi}$, and the new quantum state $\ket{\psi}$ can be prepared through $\ket{\psi} = P\ket{\phi}$. Thus, the expectation of $\ket{\phi}$ on the observable $U$ with shift operator $P$ can be simplified as follows,
\begin{equation}
\bra{\phi} P^{-1} U P \ket{\phi} =\bra{(P^{-1})^{\dagger}\phi} U \ket{P\phi}= \bra{\psi} U \ket{\psi}.
\end{equation}

\subsection{Imposition of displacement boundary conditions}
The works of Liu \emph{et al.}~\cite{Liu2021PRA} and Sato \emph{et al.}~\cite{Sato2021PRA} embraced the VQE paradigm for tackling the Poisson equation with trivial boundary conditions. In real-world contexts, it is crucial to have the capability to employ arbitrary displacement boundary conditions.

In this study, an amplitude-encoding oracle is utilized for $r$ on the right-hand side of the PDE. It is assumed that an efficient quantum routine is available for the task, thus not the primary focus of this investigation. In our approach, we apply displacement boundary conditions by selectively zeroing out the row and column associated with off-diagonal entries for the target DOF and introducing $K_{bc}$ matrix into the system and then evaluate $\bra{\psi} K_{bc} \ket{\psi}$. Notably, our proposed method employs the ``set-to-zero'' strategy, which is analogous to how conventional finite element codes manage displacement boundary conditions. This enables us to impose arbitrary displacement boundary conditions at any node within the system equations. The``set-to-zero" strategy involves the deliberate alteration of off-diagonal elements within the stiffness matrix to zero, corresponding to DOFs where boundary conditions are applied. Meanwhile, an additional correction term is introduced into the right-hand side of the original system of equations to account for this modification.

For better clarity, we elucidate this strategy in the context of a cantilever beam system with two elements. In this case, the stiffness matrix is of $6 \times 6$. The prescribed loading conditions for this system involve a unit force at the free end, with deflection and rotation at the fixed end being zero. Consequently, the resulting system of equations can be expressed as follows
\begin{eqnarray}
    K_0 \bm{u_0} = \bm{f_0}.
\end{eqnarray}

To apply the displacement boundary conditions, the off-diagonal elements of 0th and 1st DOFs are set to zero and the resulting matrix can be interpreted as adding a matrix $K_{bc}$ to the original $K_0$. In the meantime, the corresponding adjustment is added to the right-hand-side to maintain the system unchanged as
\begin{eqnarray}
    K_1 \bm{u_0} = \bm{f_1},
\end{eqnarray}
where
\begin{eqnarray}
    K_1 &=& K_0 + K_{bc}.
\end{eqnarray}
The details of this process are in Appendix~\ref{app_sec:0}.

To calculate the loss in Eq.~(\ref{eq:PiQt}) with specified boundary conditions, it is necessary to evaluate $\bra{\phi} K_{bc} \ket{\phi}$ on a quantum computer, which will be elaborated in next section.

\subsection{Evaluation of $\bra{\phi} K_{bc} \ket{\phi}$}
The expectation $\bra{\phi} K_{bc} \ket{\phi}$ can be obtained from the linear combination of the expectation of a series of symmetric matrix, each of which has a single off-diagonal element in its half matrices
\begin{eqnarray}
\bra{\phi} K_{bc} \ket{\phi} = \sum_i c_i \bra{\phi} K_{pq} \ket{\phi},
\end{eqnarray}
where $K_{pq}$ is the symmetric matrix with two non-zero off-diagonal components located at $(p, q)$ and $(q, p)$.

Generally, $K_{pq}$ cannot be linked to an observable that can be measured on a quantum computer. However, its expectation $\bra{\phi} K_{pq} \ket{\phi}$ can be manipulated to establish a connection with an observable $U$. This allows us to formulate
\begin{eqnarray}
\bra{\phi} K_{pq} \ket{\phi} = \bra{\psi} U \ket{\psi}
\end{eqnarray}
where $\ket{\psi} = T \ket{\phi}$, and $T$ is composed of a series of X and CNOT gates applied to the state $\ket{\phi}$ such that $T = \big\{ P_1 \otimes \ldots \otimes P_n : \forall i,\ P_i \in \{X, CNOT\}\big\}$, thus $T$ being unitary. $U$ is of the form $I_1^{\otimes n} \otimes X$, with $I_1= \ket{1}\bra{1}$ and $X$ the matrix representation of X gate in Eq.~(\ref{eq:pauli_gates}). The value of $\bra{\psi} I_1^{\otimes n} \otimes X \ket{\psi}$ is computed by post-processing the measurement results in computational basis. Specifically, only the probability associated with the bit string where the first $n$ bits are all `1' is considered when calculating the expectation.

The unitary transformation $T$, specifically, the sequence of X and CNOT gates, is determined by the dedicated method, which ``transfers" the non-zero terms towards the bottom-right corner. We name it ``Least Significant Bit Transformation" (LSBT) to explicitly convey its operations. It is noted that this paper focuses on the strategic implementation of imposing arbitrary boundary conditions. It is essential to highlight that a substantial portion of our research pertaining to LSBT algorithm, represents a significant and innovative contribution. Due to the depth and breadth of this content, this algorithm is presented in a separate paper currently in preparation for submission which will delve into more details. The algorithm is exemplified by a $K_{pq}$ matrix of size $4 \times 4$, which corresponds to a two-qubit system. For $ p = 0, q = 3, n = 2$, the gate sequence to convert $K_{03}$ to $I_1^{\otimes n} \otimes X $ form is $[X(0), CNOT(0, 1)]$. It is noted that $X(i)$ means a X gate is applied on the i-th qubit, while $CNOT(c, t)$ means a CONT gate is executed with c-th qubit as the control and t-th qubit as the target. The X(0) gate is applied to $K_{03}$ to transform it into $K_x$ as:
\begin{eqnarray}
    K_x = X(0)^T K_{03} X(0).
\end{eqnarray}
Then CNOT(0, 1) gate is applied to $K_x$, transforming it into $K_c$ as
\begin{eqnarray}
    K_c = CNOT(0,1)^T K_x CNOT(0, 1)
    = I_1 \otimes X = T^{T} K_{03} T,
\end{eqnarray}
where $T$ reflects the combined effect of X(0) and CNOT(0, 1). The expectation, for instance, $\bra{\psi}K_c\ket{\psi} = \bra{\phi} K_{03}\ket{\phi}$, can be evaluated on a quantum computer by applying a series of gates decomposed from $T$ to the ansatz, followed by measurements and post-processing. The detailed matrix representation and the transformation process is in Appendix~\ref{app_sec:bc}.

Therefore, the evaluation of $\bra{\phi} K_{bc} \ket{\phi}$ can be summarized as
\begin{itemize}
    \item[1.] The objective is to compute $\bra{\psi} A \ket{\psi}$ , where $A$ is an $N \times N$ symmetric matrix with zero diagonal entries and two non-zero off-diagonal entries. $N$ equals $2^n$, where $n$ is a positive integer.

    \item[2.] LSBT algorithm is proposed to derive the gate sequence of X and CNOT, and the total number of gates is of $O(n)$. The sequenced gates can combine to yield a unitary operator as $T = G_0 G_1 ... G_m$, where $G_i \in [X, CNOT]$.

    \item[3.]$A$ is reconstructed with $T$ into an observable as $T^{T} A T = I_1^{\otimes n-1} \otimes X$.

    To be specific, $(G_0 G_1 \dots G_m)^T A (G_0 G_1 \dots G_m) = I_1^{\otimes n-1} \otimes X$.

    \item[4.] Rearranging $T^{T} A T = I_1^{\otimes n-1} \otimes X$, we obtain $A = T I_1^{\otimes n-1} \otimes X T^T$, specifically, $A = (G_0 G_1 \dots G_m) I_1^{\otimes n-1} \otimes X (G_0 G_1 \dots G_m)^T = G_0 G_1 \dots G_m I_1^{\otimes n-1} \otimes X G_m^T G_{m-1}^T \dots G_1^T $.

    \item[5.] When calculating the expectation value of $A$ in the given quantum state $\psi$, the aforementioned algorithms enable $\bra{\psi} A \ket{\psi}$ = $\bra{\phi} I_1^{\otimes n-1} \otimes X \ket{\phi}$, where $\ket{\phi} = G_m^T G_{m-1}^T... G_1^T \ket{\psi}$. It is noted that $G_i \in [X, CNOT]$ being Hermitian, thus $G_i^T = G_i$ and $\ket{\phi} = G_mG_{m-1}... G_1 \ket{\psi}$ accordingly.
\end{itemize}

The operator $T$, which transforms $K_{pq}$ into the form of $I_1^{\otimes n} \otimes X$, consists of a finite number ($O(log(N))$) of X gates and CNOT gates. Since X and CNOT gates are unitary, the resultant operator $T$ is naturally unitary. The expectation $\bra{\psi} T^T K_{pq} T\ket{\psi}$ can be computed on a quantum computer using the similar methodology employed for computing terms involving the shift operator $P$ in Section~\ref{sec:4_implementation}. Up to this point, all terms in Eq.~(\ref{eq:PiQt}) with arbitrary displacement boundary conditions, can be represented as linear combinations of Pauli operators.

\subsection{State preparation}\label{sec:state_preparation}
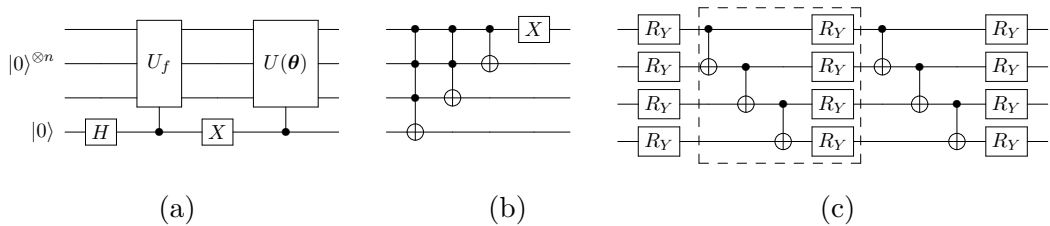
\begin{figure}
\centering
\begin{equation*}
\scalebox{0.7}{
\Qcircuit @C=1.em @R=0.4em @!R {
    \lstick{}                    & \qw        & \multigate{2}{U_f} & \qw       & \multigate{2}{U(\bm{\theta})} & \qw \\
    \lstick{\ket{0}^{\otimes n}} & \qw        & \ghost{U_f}        & \qw       & \ghost{U(\bm{\theta})}        & \qw \\
    \lstick{}                    & \qw        & \ghost{U_f}        & \qw       & \ghost{U(\bm{\theta})}        & \qw \\
    \lstick{\ket{0}}             & \gate{H}   & \ctrl{-1}          & \gate{X}  & \ctrl{-1}                & \qw                                  } \quad  \quad
}
\scalebox{0.7}{
\Qcircuit @C=1.em @R=0.4em @!R {
    & \ctrl{1} & \ctrl{1} & \ctrl{1}& \gate{X} & \qw\\
    & \ctrl{1} & \ctrl{1} & \targ   & \qw      & \qw\\
    & \ctrl{1} & \targ    & \qw     & \qw      & \qw\\
    & \targ    & \qw      & \qw     & \qw      & \qw      } \quad \quad
}   
\scalebox{0.7}{
\Qcircuit @C=1.em @R=0.4em @!R {
    & \gate{R_Y} & \ctrl{1} & \qw      & \qw      & \gate{R_Y} & \ctrl{1} & \qw      & \qw      & \gate{R_Y}  & \qw \\
    & \gate{R_Y} & \targ    & \ctrl{1} & \qw      & \gate{R_Y} & \targ    & \ctrl{1} & \qw      & \gate{R_Y}  & \qw \\
    & \gate{R_Y} & \qw      & \targ    & \ctrl{1} & \gate{R_Y} & \qw      & \targ    & \ctrl{1} & \gate{R_Y}  & \qw \\
    & \gate{R_Y} & \qw      & \qw      & \targ    & \gate{R_Y} & \qw      & \qw      & \targ    & \gate{R_Y}  & \qw
    \gategroup{1}{3}{4}{6}{.7em}{--}
}
}
\end{equation*}
\begin{equation*}
\qquad\qquad\qquad\textrm{(a)}
\qquad\qquad\qquad\qquad\qquad\textrm{(b)}
\qquad\qquad\qquad\qquad\qquad\textrm{(c)}
\qquad\qquad\qquad\qquad\qquad
\end{equation*}
\caption{(a) Quantum circuit for preparing $\ket{f, \psi(\bm{\theta})} = \left( \ket{0}\ket{f} + \ket{1}\ket{\psi(\bm{\theta})} \right) / \sqrt{2}$. (b) Quantum circuit for the shift operator $P$. (c) The ansatz for generating $\ket{\psi (\bm{\theta})}$. The dashed box outlines the repeated internal structure of the ansatz.}
\label{fig:f_psi_P_psi}
\end{figure}

Like all VQAs, the quantum state $\ket{\phi}$ is generated by applying a series of parameterized quantum circuit denoted as $U(\bm{\theta})$ to the initial state $\ket{0}^{\otimes n}$. In this study, we utilize the Qiskit RealAmplitudes template to construct the ansatz, as exemplified in Fig~.\ref{fig:f_psi_P_psi}(c). This example layout consists of alternating layers of $R_Y$ gates and CNOT gates. The RealAmplitudes circuits are specifically designed to handle real-valued amplitudes, ensuring that the amplitudes of $\ket{\phi}$ is real. This property is crucial for practical engineering applications, such as the Euler-Bernoulli beam problem in this work. Since our primary focus is on solving PDEs, we do not address the challenges related to preparing the external force term $\ket{f}$, which is another significant hurdle in quantum studies. For simplicity, we assume there exists a unitary operator, often known as the oracle, $U_f$, which encodes the amplitude of $\ket{f}$ as $\ket{f} = U_f \ket{0}^{\otimes n}$~\cite{Mottonen2005JQIC, Carrasquilla2019Nature}.

The state $\ket{f, \psi (\bm{\theta})}$ in Eq.~(\ref{eq:PiQt}) is created by utilizing an auxiliary qubit and controlled versions of both the parameterized quantum circuit $U(\bm{\theta})$ and the unitary operator $U_f$ depicted in Fig.~\ref{fig:f_psi_P_psi}(a)~\cite{Sato2021PRA}. The shift operator $P$ is implemented through a series of multi-controlled Toffoli gate together with X gate~\cite{Miller2011IEEE}, as shown in Fig.~\ref{fig:f_psi_P_psi}(b).

\subsection{Time complexity}
In this section, we analyze the time complexity of the proposed approach from various aspects, including state preparation, the quantity of quantum circuits, the number of iterations, the count of loops and amount of shots. All the indicators contribute to the optimization of parameters set $\bm{\theta}$. It's noted that the time complexity analysis focuses on the quantum computing parts, excluding the classical aspects as they reply on classical implementation and optimizers~\cite{Sato2021PRA}.

Referring to the workflow in Fig.~\ref{fig:workflow}, problems are encoded into cost functions followed by ansatz preparation and the optimal solution is progressively approached via the interplay between quantum circuits and classical optimizers. Classical optimizers fine-tune quantum circuit parameters, driving them towards convergence. In the course of this procedure, there are four crucial indicators at play to characterize the time complexity of the proposed algorithm.

The first indicator is the number of quantum circuits required to evaluated the loss in Eq.~(\ref{eq:PiQt}). The required quantity $n_q$, is $n_{q} = O(1)=19$, independent of the problem size. The second is the number of iterations. Each iteration signifies a successful step taken by the classical optimizer to determine the subsequent $\bm{\theta}$. The total number of iterations, denoted by $n_{i}$, reflects the steps to reach the optimal $\bm{\theta}$, ultimately solving the problem. The third is the count of loops. The loops indicate an internal iterative process within each "high-level" iteration, refining the parameters $\bm{\theta}$ and evaluating the cost function to enhance the solution. After running several loops within each iteration, the algorithm yields a new parameter $\bm{\theta}$ for the subsequent iteration. For simplicity, it is assumed that each iteration has a constant number of loops, i.e. $n_{l} := O(1)$. The last is the amount of the shots, which corresponds to the executions of the quantum circuit. The estimation of expectation values often demands a substantial number of shots for sampling purposes. The statistical nature of the sampling process imposes specific requirements on the number of shots to meet a predefined error threshold as $n_{s} = O(1/\varepsilon^2)$ where $\varepsilon^2$ is the mean square error of the cost function~\cite{Dekking2006}.

Consider all the indicators and assume a single shot spends the time of $t_s$. This allows us to represent the total time of the proposed algorithm as
\begin{equation}\label{eq:t}
t = t_s n_{s} n_{l} n_{i}n_q.
\end{equation}

The basic time unit $t_s$ in Eq.~(\ref{eq:t}) is affected by various factors associated with ansatz, shift operators and external forces. Among these, the Pauli measurements of all the observables in this approach increase the circuit depth by at most 3~\cite{Nielsen2010QCQI}. Therefore, we will neglect the extra time owing to these Pauli measurements and focus instead on a comprehensive examination of the other contributors in the following discussion.

The time required to prepare ansatz, $t_a$, is directly related to the circuit depth of the ansatz, $d_a$. Therefore, the time complexity of ansatz preparation can be expressed as $t_a = O(d_a)$, indicating that it scales linearly with the circuit depth of the ansatz. This time captures the duration to set up the ansatz before any specific measurements commence. Similarly, the time $t_P$ represents the duration to prepare the shift operator $P$, which is directly related to the circuit depth of the shift operator. It can be denoted as $t_P=O(n^2)$~\cite{Maslov2016PRA}. The time $t_f$ represents the duration for encoding the external force term on the right hand side. As outlined in Section~\ref {sec:state_preparation}, it is assumed that there is an oracle $U_f$ to encode the force amplitude as $\ket{f} = U_f\ket{0}^{\otimes n}$, and the specifics of its construction will not be discussed in this work.

Eq.~(\ref{eq:PiQt}) can yield the values of $\bra{f,\phi(\bm{\theta})} X \otimes I^{\otimes n} \ket{ f, \phi(\bm{\theta})}$ and $\bra{\phi(\bm{\theta})} K \ket{\phi(\bm{\theta})}$. To be specific, $\bra{f, \phi(\bm{\theta})} X \otimes I^{\otimes n} \ket{ f, \phi(\bm{\theta})}$ involves the ansatz preparation and the force amplitude encoding, which takes $t_{a} + t_f$. $\bra{\phi(\bm{\theta})} K \ket{\phi(\bm{\theta})}$ involves the ansatz preparation for the six terms in Eq.~(\ref{eq:Ke}), these same six terms coupled with the shift operator $P$ and the additional six terms coupled with the Hermitian $I_0$ as discussed in Section \ref{sec:K_exp}, denoted as $16t_a+11t_P$. According to the proposed LSBT algorithm, the boundary condition term $\bra{\phi(\bm{\theta})} K_{bc} \ket{\phi(\bm{\theta})}$ signifies each prescribed boundary condition at any single node entails the evaluation of five terms, expressed as $5t_{bc}+5t_a$. It exhibits the time complexity $t_{bc} = O(n)$. Assuming there are $m$ prescribed boundary values, the total time for a shot can be calculated as
\begin{equation}
    t_s = \underbrace{t_{a} +t_f}_{f,\psi(\theta)} + \underbrace{16t_a+11 t_P}_K + \underbrace{5 m \cdot (t_{bc}+t_a)}_{K_{bc}} =(17+5m)t_a+11t_P+5m\cdot t_{bc} +t_f = O(d_{a} + n^2 + n + t_f).
\end{equation}
Hence, the total time complexity is
\begin{equation}
    O\left( \frac{ n_q n_{i} n_{l}(d_{a} + n^2 + t_f)}{\varepsilon^2}\right).
\end{equation}

While we tackle increasingly intricate problems, our algorithm upholds a consistent order of time complexity as the novel VQA approach for solving Poisson equation with $O\left({t_i(d_a+d_e+n^2)nd_a}/{\varepsilon^2}\right)$ \cite{ Sato2021PRA}. This similarity arises because ${\varepsilon^2}$ as the denominator plays a dominant role in both equations. Moreover, our algorithm presents an advantage over another efficient VQA for Poisson equation with $ O\left({t_i(d_a+d_e)n^2d_a}/{\varepsilon^2}\right)$ \cite{Liu2021PRA}, wherein the subscript $e$ indicates amplitude encoding.
\section{Results and discussion}\label{sec:5_results}
\begin{figure} 
\centering
\includegraphics[width=0.7\textwidth]{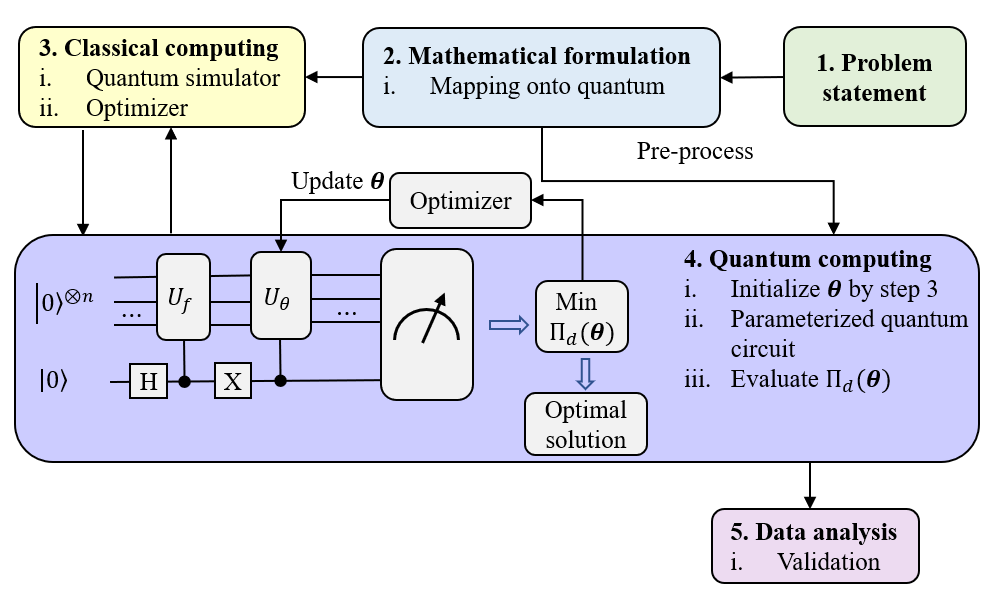}
\caption{Workflow illustrates the proposed method that integrates quantum and classical computing approaches, where the objective function $\Pi_d(\bm{\theta})$ is defined as a function of the quantum circuit parameters $\bm{\theta}$. The number of parameters $\bm{\theta}$, denoted by $i$, varies according to the ansatz structure.}
\label{fig:workflow}
\end{figure}

This section depicts the results of the numerical experiments to demonstrate the predictive capability of the proposed algorithm across diverse boundary conditions. Qiskit ver 0.37.1 was utilized with Python 3.10 to implement the proposed method. The calculations were performed utilizing the \textit{statevector simulator} backend. It models the behavior of a quantum circuit in terms of its underlying quantum state and returns the exact state vectors of the quantum states. The Broyden–Fletcher–Goldfarb–Shanno (BFGS)~\citep{Broyden1970IMA, Fletcher1970TCJ, Goldfarb1970MOC,Shanno1970MOC} optimizer was utilized to update parameters $\bm{\theta}$. The workflow for the proposed algorithm are summarized in the flowchart in Fig.~\ref{fig:workflow}. After problem identification, the classical formulation is translated into a quantum formulation. The simulator is utilized to replicate the behavior of a real quantum device, generating predictions for the objective function $\Pi_d(\bm{\theta})$, which involves a series of quantum circuit parameters denoted as $\bm{\theta}(i)$. Subsequently, the predictions are optimized using a classical optimizer by iteratively updating the values of $\bm{\theta}(i)$ and ultimately leading to an optimal solution.

\begin{figure}
    \begin{equation*}
    \scalebox{0.7}{
    \Qcircuit @C=.5em @R=.5em {
        \lstick{q_0} & \gate{R_Y({\theta_0})} & \ctrl{1} & \qw      & \qw      & \qw      & \gate{R_Y({\theta_5})} & \ctrl{1} & \qw       & \qw      & \qw      & \gate{R_Y({\theta_{10})}} & \ctrl{1} & \qw       & \qw      & \qw      & \gate{R_Y({\theta_{15}})} & \ctrl{1} & \qw       & \qw      & \qw      & \gate{R_Y({\theta_{20}})} & \ctrl{1} & \qw       & \qw      & \qw      & \gate{R_Y({\theta_{25}})} &\qw \\
        \lstick{q_1} & \gate{R_Y({\theta_1})} & \targ    & \ctrl{1} & \qw      & \qw      & \gate{R_Y({\theta_6})} & \targ    & \ctrl{1}  & \qw      & \qw      & \gate{R_Y({\theta_{11}})} & \targ    & \ctrl{1}  & \qw      & \qw      & \gate{R_Y({\theta_{16}})} & \targ    & \ctrl{1}  & \qw      & \qw      & \gate{R_Y({\theta_{21}})} & \targ    & \ctrl{1}  & \qw      & \qw      & \gate{R_Y({\theta_{26}})} &\qw \\
        \lstick{q_2} & \gate{R_Y({\theta_2})} & \qw      & \targ    & \ctrl{1} & \qw      & \gate{R_Y({\theta_7})} & \qw      & \targ     & \ctrl{1} & \qw      & \gate{R_Y({\theta_{12}})} & \qw      & \targ     & \ctrl{1} & \qw      & \gate{R_Y({\theta_{17}})} & \qw      & \targ     & \ctrl{1} & \qw      & \gate{R_Y({\theta_{22}})} & \qw      & \targ     & \ctrl{1} & \qw      & \gate{R_Y({\theta_{27}})} &\qw \\
        \lstick{q_3} & \gate{R_Y({\theta_3})} & \qw      & \qw      & \targ    & \ctrl{1} & \gate{R_Y({\theta_8})} & \qw      & \qw       & \targ    & \ctrl{1} & \gate{R_Y({\theta_{13}})} & \qw      & \qw       & \targ    & \ctrl{1} & \gate{R_Y({\theta_{18}})} & \qw      & \qw       & \targ    & \ctrl{1} & \gate{R_Y({\theta_{23}})} & \qw      & \qw       & \targ    & \ctrl{1} & \gate{R_Y({\theta_{28}})} &\qw \\
        \lstick{q_4} & \gate{R_Y({\theta_4})} & \qw      & \qw      & \qw      & \targ    & \gate{R_Y({\theta_9})} & \qw      & \qw       & \qw      & \targ    & \gate{R_Y({\theta_{14}})} & \qw      & \qw       & \qw      & \targ    & \gate{R_Y({\theta_{19}})} & \qw      & \qw       & \qw      & \targ    & \gate{R_Y({\theta_{24}})} & \qw      & \qw       & \qw      & \targ    & \gate{R_Y({\theta_{29}})} &\qw
        \gategroup{1}{3}{5}{7}{.7em}{--}
    }
    }
    \end{equation*}
    \caption{Circuit diagram showcases the ansatz with linear entanglement employed in this work. The ansatz consists of five layers acting on an five-qubit register, $q_i$, initialized to the computational basis state $\ket{0} ^{\otimes 5}$ where $q_i=\ket{0}$. Each layer comprises CNOT gates arranged in a linear entanglement structure followed by single-qubit rotation gates $R_Y$, applied to each qubit, as shown in the dashed box.}
    \label{fig:ansatz}
\end{figure}
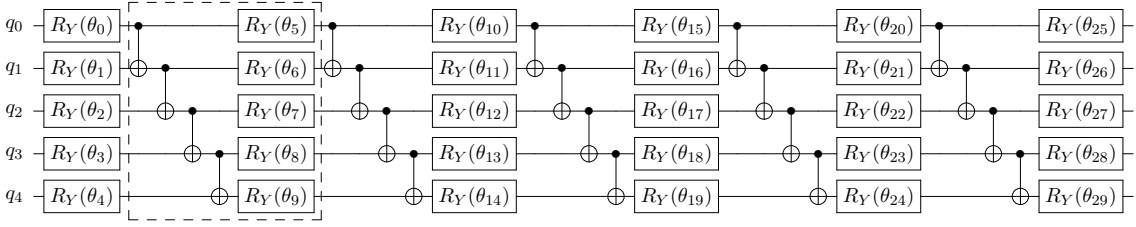

The proposed model was tested on an Euler-Bernoulli beam with four distinct boundary conditions, including the beam with a periodic boundary condition (PBC), the cantilever beam, the simply-supported beam (SSB), and the fixed-fixed beam (FFB). The boundary conditions detailed in Fig.~\ref{fig:Results}(SCHEMATIC) encompass various scenarios including periodic, free, fixed, and pinned ends. This comprehensive selection is designed to bolster the demonstration's completeness and reliability. It not only highlights the model's adaptability to diverse boundary conditions but also validates its predictive accuracy in different contexts. The following part presents the simulated results obtained using the four representative boundary conditions, while maintaining a consistent configuration of five qubits, five ansatz layers as illustrated in Fig. \ref{fig:ansatz}, and a beam length of 10 $m$ with Young's modulus of 1000 $Pa$ and second moment of inertia of 1.0 $m^2$.

\subsection{Convergence study}
\begin{figure}[hbt!]
    \centering
    \includegraphics[width=\textwidth]{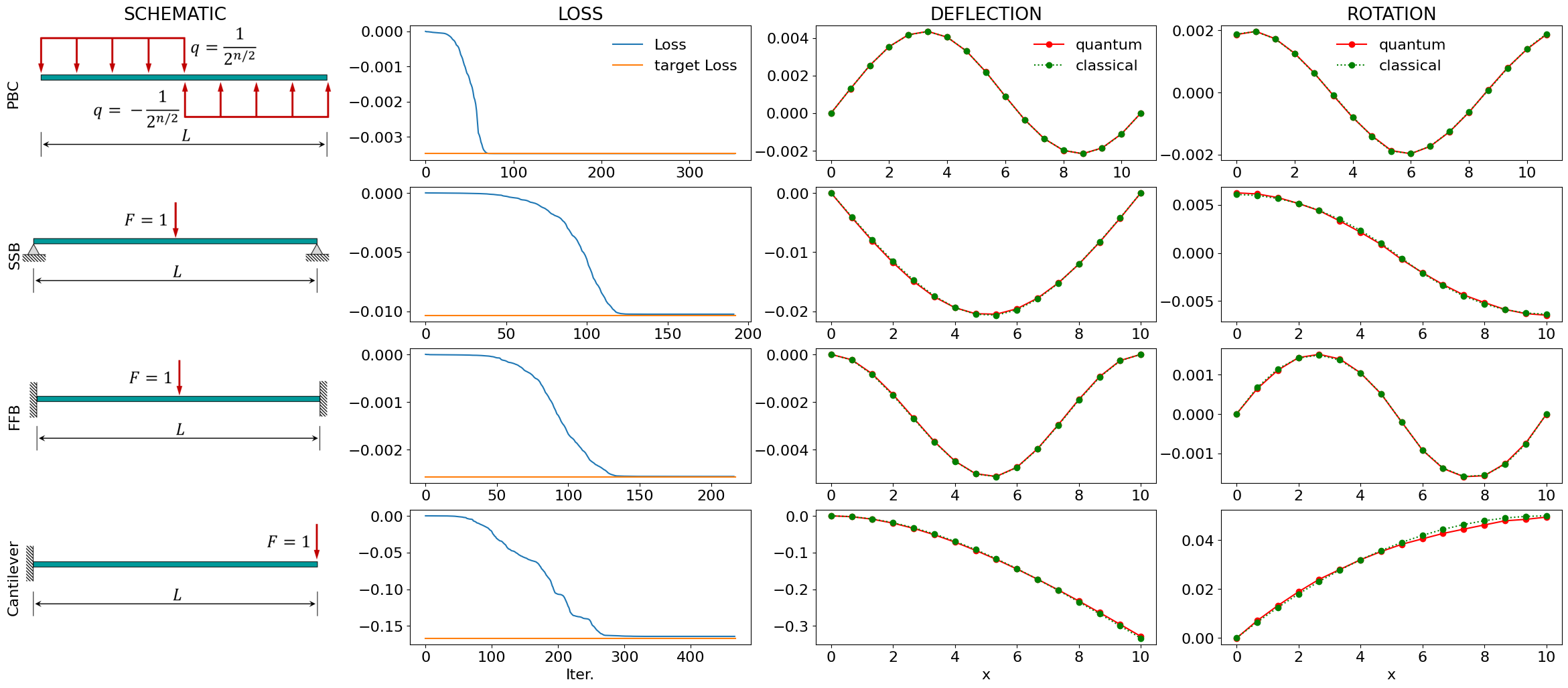}
    \caption{Schematics, loss convergence history, deflection (unit: $m$) and rotation (unit: $rad$) of the four representative boundary problems. The first (SCHEMATIC) column depicts the schematics of the four representative boundary conditions, namely PBC, SSB, FFB, and Cantilever. The second (LOSS) column is the temporal evolution of the objective functions throughout the simulation iterations, illustrating its convergence towards the target loss (the yellow solid line). The third (DEFLECTION) and the fourth (ROTATION) column display the simulated beam profiles for the deflection and rotation. The red solid line corresponds to quantum computing approach while the green dotted line represents results from classical methods.}
    \label{fig:Results}
\end{figure}

Figs.~\ref{fig:Results}(LOSS) depict the temporal evolution of the loss function values, converging towards the target values. Each iteration refines the quantum circuit parameters through a classical optimizer. The exceptional applicability, robustness, and efficiency of our methodologies are exemplified by their ability to address diverse boundary conditions. Additionally, the established classical optimizer's proficiency in tailoring parameters based on the feedback from the objective function enables quantum algorithms like VQA to adjust to the noise profile present in real quantum devices \citep{Ravi2022}. This capability significantly reduces the detrimental effects of noise, enhancing the performance of quantum algorithms. Such insights are invaluable for researchers engaged with genuine quantum devices.

Figs.~\ref{fig:Results}(LOSS) demonstrate the convergence of the predicted loss value towards the desired target loss. The average number of iterations across four boundary conditions is approximately 174. In particular, when considering the PBC, the iteration count is 77, aligning well with the conclusions of Pellow \emph{et. al}~\cite{Pellow2021QIP}, which highlighted the exceptional performance and fast convergence rate achieved when employing the similar setup and the study conducted by Wierchs \emph{et. al}~\cite{Wierichs2020PRR} which demonstrated fast and reliable outcomes. In a nutshell, our proposed model can arrive at the same conclusion as the well-established algorithms in terms of speed and accuracy.

\begin{figure}
\centering
\begin{tikzpicture}[scale=1.]
\definecolor{newred}{rgb}{1.00,0.70,0.70}
\begin{axis}[width=.6\textwidth, height=.3\textheight,
    ybar, bar width=0.5cm, enlarge x limits=0.15,
    legend style={at={(0.5,-0.15)},
    anchor=north,legend columns=-1},
    ytick={0,50,...,350},
    ymax=350,ymin=0, axis y line*=left,
    yticklabels={0,50,...,350},
    y label style={at={(0.0,0.5)}},
    ylabel={Iteration times},
    symbolic x coords={PBC, SSB, FFB,Cantilever}, xtick=data,
    xtick style={draw=none},bar shift=-9,]
    \addplot coordinates {(PBC,77) (SSB,134) (FFB,147)(Cantilever,339)};
    \addlegendentry{Iteration}
    \addlegendimage{color=red,fill=newred}
    \addlegendentry{Accuracy}
\end{axis}
\begin{axis}[width=.6\textwidth, height=.3\textheight,
    ybar, bar width=0.5cm, enlarge x limits=0.15, axis y line*=right,
    legend style={at={(0.5,-0.15)},
    anchor=north,legend columns=-1},
    ytick={93,94,95,96,97,98,99,100},
    ymax=100,ymin=93,
    ytick={93,94,95,96,97,98,99,100},
    y label style={at={(0,0.5)}},
    ylabel={Acccuracy $\%$}, ylabel near ticks,
    symbolic x coords={PBC, SSB, FFB, Cantilever}, xtick=data, xticklabels={},xtick style={draw=none}, bar shift=9,]
    \addplot[color=red,fill=newred] coordinates {(PBC,99.87896734) (SSB,99.1708881) (FFB,99.74898817) (Cantilever,98.5492029)};
\end{axis}
\end{tikzpicture}
\caption{The bar charts depict the relationship between iteration times (primary axis) based on Fig. \ref{fig:Results} and accuracy level (secondary axis) upon convergence. The accuracy level is quantified by the variation between the target loss and the corresponding predicted result at convergence, scaled by a factor of 100.}
\label{fig:Con_accuracy}
\end{figure}
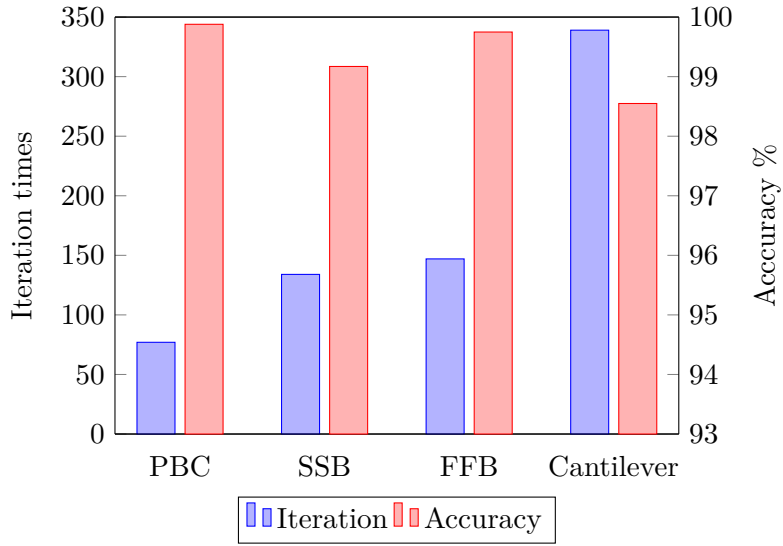

Upon further examination of the convergence process, we observe distinct patterns in the iteration count and the accuracy level characterized by the ratio between the target and predicted loss values. Fig.~\ref{fig:Results}(Cantilever-LOSS) demonstrates that despite converged, a slight discrepancy remains between the target and predicted loss. Conversely, Fig. \ref{fig:Results}(PBC-LOSS) showcases that the fewest iterations corresponds to a seamless alignment of the target and predicted loss upon convergence. The SSB and FFB exhibit intermediate gaps and iteration counts. The findings have been compactly summarized in Fig. \ref{fig:Con_accuracy}. It characterizes the accuracy level using the indicator of relative error adopted in \eg Uvarov~\emph{et. al}~\cite{Uvarov2020PRB} or Sato~\emph{et. al} ~\cite{Sato2023arXiv} based on
\begin{equation*}
\label{eq:accuracy}
    \textrm{Accuracy}~\% = 100 \times(1- \frac{\left | \mathcal{O}_\textrm{t} - \mathcal{O}_\textrm{p} \right |}{\left | \mathcal{O}_\textrm{t} \right | }),
\end{equation*}
where $\mathcal{O}_\textrm{t}$ and $\mathcal{O}_\textrm{p}$ represent the target and predicted loss, respectively. Here ${\left |\mathcal{O}_\textrm{t} - \mathcal{O}_\textrm{p} \right |}/{\left | \mathcal{O}_\textrm{t} \right | }$ is called the relative error, typically not multiplied by 100 $\%$.

The accuracy levels consistently exceed 98.5$\%$, with the PBC case reaching up to 99.9$\%$ and averaging at 99.3$\%$. In other words, the relative error is less than 0.015 among all the cases shown in Fig.\ref{fig:Relative_error} in Appendix~\ref{app_sec:supp_plots}. This data adds significant weight to the capability and reliability of the proposed method. The figure also includes the total iteration counts upon convergence. The integration of the two parameters in Fig.~\ref{fig:Con_accuracy} clearly displays an inverse relationship between prediction accuracy and iterations. Specifically, the prediction accuracy tends to decrease as the total number of iterations increase. This is because more iterations result from the slower or challenging convergence to the target, potentially leading to reduced accuracy. This relationship allows for an early-stage assessment of prediction quality. An abundance of iterations suggests that the final prediction is more likely to fall short of the target.

\subsection{Prediction Accuracy}
In the overall evaluation, the proposed model exhibits strong consistency with the classical solutions across the four distinct boundary conditions, as illustrated in Figs. \ref{fig:Results}, \ref{fig:Con_accuracy}, and \ref{fig:Relative_error}. Fig. \ref{fig:Con_accuracy} and \ref{fig:Relative_error} showcasing the prediction accuracy in terms of objective values, offer intuitive and convincing evidence. Fig. \ref{fig:Results} represent the objective values, deflection and rotation profiles of a simulated beam, respectively. While a slight discrepancy is observed in the cantilever case, all cases exhibit similar shapes and closely match the classical solutions, indicating a high level of accuracy. To quantify the differences, Root Mean Square Error (RMSE) is employed to evaluate disparities between predicted and actual values. As shown in Fig. \ref{fig:RMSE} in Appendix~\ref{app_sec:supp_plots}, except for the cantilever beam, all cases demonstrate RMSE values below 1.3$\times 10^{-4}$. In particular, the PBC and FFB cases necessitate zooming in for visualization due to their extremely low RMSE values. Despite the RMSE values for the cantilever case ranging from approximately 1.0$\times 10^{-3}$ to 2.4$\times 10^{-3}$, it is important to consider the scope of the target values. The loss, deflection and rotation values span from -0.15 to 0, 0 to 0.3, and 0 to 0.05, respectively. Given the data ranges, the average deviation between predictions and true values is only on the order of $10^{-3}$, suggesting a reasonable level of accuracy, as evidenced by Fig. \ref{fig:Results}(Cantilever).

To further assess the quality of the solution, normalization is performed on RMSE using the max/min value, $v_\textrm{max}$ and $v_\textrm{min}$ from the classical results, as given by
\begin{equation*}
\textrm{RMSE}_\textrm{n} ~\%= \frac{\textrm{RMSE}} {v_\textrm{max}-v_\textrm{min}}\times 100.
\label{eq:Norma_RMSE}
\end{equation*}

\begin{figure}
\centering
\begin{tikzpicture}[scale=1.]
\definecolor{newblue}{rgb}{0.70,0.70,1.00}
\definecolor{newred}{rgb}{1.00,0.70,0.70}
\definecolor{newbrown}{rgb}{0.925,0.850,0.775}
\definecolor{newbrownborder}{rgb}{0.45,0.30,0.15}
\begin{axis}[width=.7\textwidth, height=.3\textheight,
    ybar, bar width=0.5cm, enlarge x limits=0.15, axis y line*=right,
    legend style={at={(0.5,-0.15)},
    anchor=north,legend columns=-1},
    ytick={0.99975,0.9998,0.99985,0.9999,0.99995,1.0},
    ymax=1,ymin=0.99975,
    yticklabels={0.99975,0.9998,0.99985,0.9999,0.99995,1.0},
    y label style={at={(-0.05,0.5)}},
    ylabel={Fidelity},ylabel near ticks,
    symbolic x coords={PBC, SSB, FFB, Cantilever}, xtick=data,
    xticklabels={}, xtick style={draw=none}, bar shift =17]
    \addplot [color=newbrownborder,fill=newbrown] coordinates {(PBC,0.99997560219049) (SSB,0.999859801025005) (FFB,0.999945500358205)(Cantilever,0.99986137574219) };
\end{axis}
\begin{axis}[width=.7\textwidth, height=.3\textheight,
    ybar, bar width=0.5cm, enlarge x limits=0.15,
    legend style={at={(0.5,-0.15)},axis y line*=left,
    anchor=north,legend columns=-1},
    ytick={0,0.5,...,2.5},
    ymax=2.5,ymin=0, axis on top,
    yticklabels={0,0.5,...,2.5},
    y label style={at={(0.0,0.5)}},
    ylabel={Normalized RMSE $\%$},
    symbolic x coords={PBC, SSB, FFB, Cantilever}, xtick=data,
    xtick style={draw=none},
    ]
\addplot[bar shift=-20,blue,fill=newblue] coordinates {(PBC,0.082960048) (SSB,0.629500497)(FFB,0.382331971)(Cantilever,0.687503912)};
\addplot[bar shift=-2,red,fill=newred] coordinates {(PBC,0.328906018) (SSB,0.935971681)(FFB,0.570123524)(Cantilever,2.119260172)};
\legend{Deflection, Rotation}
\draw [blue, thick,dotted,](axis cs:PBC, 0.445574107)--(axis cs:Cantilever, 0.445574107);
\draw [red, thick,dotted,](axis cs:PBC, 0.988565349)--(axis cs:Cantilever, 0.988565349);
\node [blue, anchor=west,xshift=-20pt,] at (axis cs:PBC, 0.6) {Deflection avg.};
\node [red, anchor=west,xshift=-20pt,] at (axis cs:PBC,1.1) {Rotation avg.};
\addlegendimage{color=newbrownborder,fill=newbrown}
\addlegendentry{Fidelity}
\end{axis}
\end{tikzpicture}
\caption{Accuracy indicators: 1) Quantitative analysis of normalized RMSE (primary axis) scaled by a factor of 100 for deflection and rotation across four boundary conditions compared with the classical solutions. Average Values are indicated by dashed lines. 2) Fidelity analysis (secondary axis) across four boundary conditions with respect to the classical solutions. }
\label{fig:NRMSE_FD}
\end{figure}
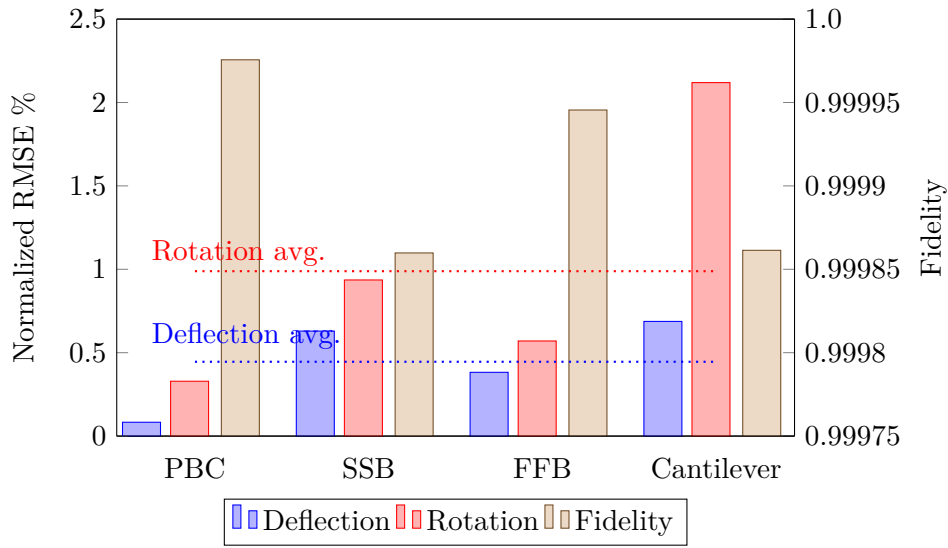

Consequently, the resulting Fig.~\ref{fig:NRMSE_FD} illustrates the normalized RMSE values for deflection and rotation, alongside the corresponding average values. Clearly, the majority of normalized RMSE values are below 1.0 $\%$, indicating excellent prediction accuracy and coinciding with the observations from the RMSE analysis. The normalized RMSE for cantilever beam rotation is approximately 2.0 $\%$, which is relatively higher compared to other cases. However, considering its small absolute value, this performance is still acceptable. The deviation observed in the cantilever case can be attributed to the additional complexity introduced by its boundary conditions, specifically one fixed and one free ends. The presence of asymmetric boundary conditions complicates the formulation of matrices and overall expressions in the analysis. The overall performance of the proposed method is further evaluated by computing average normalized RMSE $\%$ values, with deflection below 0.5 $\%$ and rotation below 1.0 $\%$.

In conjunction with RMSE metric, the fidelity of the proposed method is evaluated as a means to assess its comprehensive performance, as described by
\begin{equation*}
    \mathcal{F}= |\langle\psi_\theta|\psi_\phi\rangle|^2,
    \label{eq:fidelity}
\end{equation*}
where $|\psi_\theta\rangle$ and $|\psi_\phi\rangle$ are the state vectors from quantum simulations and classical solutions, respectively. This metric quantifies the proximity between the simulated and the corresponding reference states. A fidelity of unity signifies an identity between the two states while a value of zero indicates their orthogonality. The latter portion of Fig. \ref{fig:NRMSE_FD} illustrates that the fidelity values across all the four cases approach unity, thereby manifesting a conspicuous demonstration of the great predictive accuracy declared by the RMSE metric.

These results highlight the superior capabilities of the proposed algorithm, showcasing their great potential for accommodating diverse boundary conditions. In addition, the good prediction accuracy suggests that the proposed model can perform on par with classical methods, effectively leveraging the advantages of quantum computing while retaining the strengths of classical approaches. This integration holds substantial significance when tackling large-scale problems that may prove intractable using classical methods alone.

\begin{figure}
\centering
\hspace{-1.4cm}
\begin{tikzpicture}
\begin{axis}[
    ybar, bar width=0.5cm,
    width=.6\textwidth, height=.3\textheight,
    ylabel={RMSE},ymin=0, ymax=2.5e-3, ytick={0,0.5e-3,...,2.5e-3}, y label style={at={(0.,0.5)}},
    symbolic x coords={PBC,SSB,FFB,Cantilever},
    xtick=data, enlarge x limits={abs=30pt},
    xtick style={draw=none},
    legend style={at={(0.5,-0.15)},
    anchor=north,legend columns=-1},
]
\addplot coordinates{(PBC,0.00421e-3) (SSB,0.0856e-3) (FFB,0.00645e-3)(Cantilever,2.418e-3)};
\addlegendentry{Objective}
\addplot coordinates{(PBC,0.005395427e-3) (SSB,0.129982959e-3) (FFB,0.019648822e-3)(Cantilever,2.291679902e-3)};
\addlegendentry{Deflection}
\addplot coordinates{(PBC,0.012920659e-3) (SSB,0.116476572e-3) (FFB,0.017568268e-3)(Cantilever,1.059629743e-3)};
\addlegendentry{Rotation}
\end{axis}
\begin{axis}[
    ybar, bar width=0.22cm,
    width=.25\textwidth, height=.14\textheight, at={(2.7cm,4.9cm)},
    ymin=0, ymax=0.014e-3, ytick={0,0.007e-3,0.014e-3},
    symbolic x coords={PBC_o,PBC_d,PBC_r},xtick=data,
    xticklabels={},
    enlarge x limits={abs=20pt},
    xtick style={draw=none},
    anchor=north,legend columns=-1,
]
\definecolor{newblue}{rgb}{0.70,0.70,1.00} 
\definecolor{newred}{rgb}{1.00,0.70,0.70} 
\definecolor{newbrown}{rgb}{0.925,0.850,0.775} 
\definecolor{newbrownborder}{rgb}{0.45,0.30,0.15}
\addplot[color=blue,fill=newblue] coordinates{(PBC_o,0.00421e-3)};
\addplot[color=red,fill=newred] coordinates{(PBC_d,0.005395427e-3)};
\addplot[color=newbrownborder, fill=newbrown] coordinates{(PBC_r,0.012920659e-3)};
\node[above=12pt,xshift=-18pt] at (axis cs:PBC_d,0.005395427e-3) {PBC};
\end{axis}
\begin{axis}[
    ybar, bar width=0.22cm,
    width=.25\textwidth, height=.14\textheight, at={(2.7cm,2.6cm)},
    ymin=0, ymax=0.021e-3, ytick={0,0.01e-3,0.02e-3},
    symbolic x coords={FFB_o,FFB_d,FFB_r},xtick=data,
    xticklabels={}, xmin=FFB_o, xmax=FFB_r,
    enlarge x limits={abs=20pt},
    xtick style={draw=none},
    anchor=north,legend columns=-1,
]
\definecolor{newblue}{rgb}{0.70,0.70,1.00} 
\definecolor{newred}{rgb}{1.00,0.70,0.70} 
\definecolor{newbrown}{rgb}{0.925,0.850,0.775} 
\definecolor{newbrownborder}{rgb}{0.45,0.30,0.15} 
\addplot[color=blue,fill=newblue] coordinates{(FFB_o,0.00645e-3)};
\addplot[color=red,fill=newred] coordinates{(FFB_d,0.019648822e-3)};
\addplot[color=newbrownborder, fill=newbrown] coordinates{(FFB_r,0.017568268e-3)};
\node[below=0pt,xshift=-18pt] at (axis cs:FFB_d,0.019648822e-3) {FFB};
\end{axis}
\end{tikzpicture}
\caption{Analysis of RMSE for objective, deflection, and rotation under four
boundary conditions compared to the classical solutions. The top left section
zooms in on PBC and FFB due to limitations in visualization.}
\label{fig:RMSE}
\end{figure}
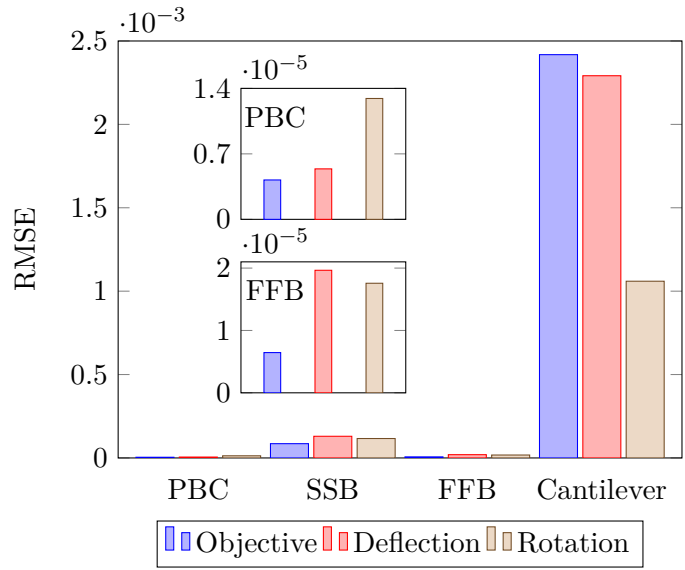

\section{conclusions}\label{sec:6_conclusion}
This paper has presented a successful application of VQAs within the FEM framework, specifically solving the fourth-order PDE Euler-Bernoulli problems.
This study extends the conventional focus on second-order PDEs to fourth-order PDEs, thereby expanding the potential application of VQAs beyond the scope of Poisson equations.
It also enables the incorporation of material characteristics and geometric flexibility.
Moreover, a systematic strategy has been proposed for imposing general boundary conditions within the quantum framework for discretized FEM equations, which has never been reported by others, and can be naturally extended to other methods such as FDM or FVM.
The novel LSBT algorithm enhances the quantum compatibility and feasibility of solving general PDE problems.
It enables us to handle matrices with structures that are not readily implementable in quantum computing and its applicability extends beyond this context.
Additionally, this study demonstrates the scalability of the method since the stiffness matrix can be Pauli decomposed into constant terms $\mathcal{O}(1)$.
The validity and accuracy of the proposed algorithm have been proved and demonstrated through four case studies, which showcased good accuracy compared with the classical solutions.
This research pioneers the expansion of VQAs to fourth-order PDEs within the FEM framework.
The ability to handle general boundary conditions, scalability, and competitive predictive capability makes this approach a promising tool for tackling a wide range of practical engineering problems.

\begin{acknowledgments}
This research is supported by the National Research Foundation, Singapore and A*STAR under its Quantum Engineering Programme (NRF2021-QEP2-02-P04).
\end{acknowledgments}

\appendix\label{sec:7_app}
\setcounter{figure}{0}
\renewcommand{\thefigure}{A\arabic{figure}}
\section{Element stiffness matrix}
\label{app_sec:K}
The element stiffness matrix $K_e$ of the Euler-Bernoulli beam is,
\begin{eqnarray}
    K_e =
    \begin{bmatrix}
        \frac{ 12EI}{l_e^3} & \frac{  6EI}{l_e^2} & \frac{-12EI}{l_e^3} & \frac{  6EI}{l_e^2} \\
        \frac{  6EI}{l_e^2} & \frac{  4EI}{l_e  } & \frac{- 6EI}{l_e^2} & \frac{  2EI}{l_e  } \\
        \frac{-12EI}{l_e^3} & \frac{- 6EI}{l_e^2} & \frac{ 12EI}{l_e^3} & \frac{- 6EI}{l_e^2} \\
        \frac{  6EI}{l_e^2} & \frac{  2EI}{l_e  } & \frac{- 6EI}{l_e^2} & \frac{  4EI}{l_e  } \\
    \end{bmatrix}.
\end{eqnarray}

Using $E=1$, $I=1$, and $l_e=1$, $K_e$ can be computed and decomposed into
\begin{eqnarray}
    K_e =
    \begin{bmatrix}
        12 & 6& -12& 6\cr
        6 & 4& -6& 2\cr
        -12 & -6& 12& -6\cr
        6 & 2& -6& 4\cr
    \end{bmatrix}
      &=& 8 I \otimes I +4 I \otimes Z -5 X \otimes I -7 X \otimes Z -6 Y \otimes Y +6 Z \otimes X.
\end{eqnarray}

The matrix $K$ in Eq.~(\ref{eq:PiQt}) is,
\begin{equation}
    K =
\begin{bmatrix}
\begin{array}{ccccccccccccc}
12  & 6    & -12  & 6 &      &  &      &      &   &    &      &      &\\
6   & 4    & -6   & 2 &      &  &      &      &   &    &      &      &\\
-12 &-6    & 24   & 0 & -12  &6 &      &      &   &    &      &      &\\
6   & 2    & 0    & 8 &  -6  &2 &      &      &   &    &      &      &\\
    &      & -12  & -6&      &  &      &\ddots&   &    &      &      &\\
    &      &  6   & 2 &      &  &      &      &   &    &      &      &\\
    &      &      &   &      &  &\ddots&      &   &    &      &      &\\
    &      &      &   &\ddots&  &      &      &   &-12 &6     &      &\\
    &      &      &   &      &  &      &      &   &-6  &2     &      &\\
    &      &      &   &      &  &      &-12   &-6 &24  &0     &-12   & 6\\
    &      &      &   &      &  &      &6     &2  & 0  & 8    & -6   & 2\\
    &      &      &   &      &  &      &      &   &-12 & -6   & 12   & -6\\
    &      &      &   &      &  &      &      &   & 6  & 2    & -6   & 4
\end{array}
\end{bmatrix}.
\end{equation}

 It can be partitioned into three interconnected components associated with $K_e$ as
\begin{eqnarray}
    K &=& K_0 + K_1 + K_2,
\end{eqnarray}
where
\begin{eqnarray}\label{eq:K0}
K_0 = \begin{bmatrix}
12 & 6 & -12& 6 &      &   &   &    &  \\
6  & 4 & -6 & 2 &      &   &   &    &  \\
-12& -6& 12 & -6&      &   &   &    &  \\
6  & 2 & -6 & 4 &      &   &   &    &  \\
   &   &    &   &\ddots&   &   &    &   \\
   &   &    &   &      & 12& 6 & -12& 6 \\
   &   &    &   &      & 6 & 4 & -6 & 2 \\
   &   &    &   &      &-12& -6& 12 & -6 \\
   &   &    &   &      & 6 & 2 & -6 & 4
\end{bmatrix},
\end{eqnarray}

\begin{eqnarray}\label{eq:K1}
    K_1 =
\begin{bmatrix}
\begin{array}{cccccccccccccc}
12 & -6&   &  &      &    &      &    &  &    &      &-12&-6\\
-6 &  4&   &  &      &    &      &    &  &    &      & 6 &2\\
   &   & 12& 6& -12  & 6  &      &    &  &    &      &   &\\
   &   & 6 & 4& -6   & 2  &      &    &  &    &      &   &\\
   &   &-12& 6& 12   & -6 &      &    &  &    &      &   &\\
   &   & 6 & 2& -6   &4   &      &    &  &    &      &   &\\
   &   &   &  &      &    &\ddots&    &  &    &      &   &\\
   &   &   &  &      &    &      &12  &6 &-12 &6     &   &\\
   &   &   &  &      &    &      &6   &4 &-6  & 2    &   &\\
   &   &   &  &      &    &      &-12 &6 &12  &-6    &   &\\
   &   &   &  &      &    &      &6   &2 &-6  &4     &   &\\
-12& 6 &   &  &      &    &      &    &  &    &      & 12&6\\
-6 & 2 &   &  &      &    &      &    &  &    &      & 6 &4
\end{array}
\end{bmatrix} ,
\end{eqnarray}

\begin{eqnarray}\label{eq:K2}
    K_2 =
\begin{bmatrix}
12 & -6& 0 &  &\cdots&  &0  &-12& -6 \\
-6 & 4 & 0 &  &\cdots&  &0  & 6 & 2 \\
0  & 0 & 0 &  &\cdots&  &  0& 0 & 0 \\
\vdots&\vdots&\vdots& &\ddots& &\vdots&\vdots&\vdots\\
0  & 0 &0  &  &\cdots&  & 0 & 0 & 0  \\
-12& 6 & 0 &  &\cdots&  & 0 & 12& 6 \\
-6 & 2 & 0 &  &\cdots&  & 0 & 6 & 4
\end{bmatrix}.
\end{eqnarray}

\section{The ``Set-to-zero" strategy}
\label{app_sec:0}
Consequently, the resulting system of equations can be expressed as follows
\begin{eqnarray}
    K_0 \bm{u_0} = \bm{f_0},
\end{eqnarray}
Following Eq.~(\ref{eq:K012}), we obtain
\begin{eqnarray}
    \begin{bmatrix}
    12 & 6 & -12 & 6 & 0 & 0 \\
    6 & 4 & -6 & 2 & 0 & 0 \\
    -12 & -6 & 24 & 0 & -12 & 6 \\
    6 & 2 & 0 & 8 & 6 & 2 \\
    0 & 0 & -12 & -6 & 12 & -6 \\
    0 & 0 & 6 & 2 & -6 & 4
    \end{bmatrix}
    \begin{bmatrix}
    0\\ 0\\ v3\\ v4\\ v5\\ v6
    \end{bmatrix}
    =
    \begin{bmatrix}
    0\\ 0\\ 0\\ 0\\ 1\\ 0
    \end{bmatrix},
\end{eqnarray}
To apply the displacement boundary conditions, the off-diagonal elements of 0th and 1st DOFs are set to zero as
\begin{eqnarray}
    \begin{bmatrix}
    12 & 0 & 0 & 0 & 0 & 0 \\
    0 & 4 & 0 & 0 & 0 & 0 \\
    0 & 0 & 24 & 0 & -12 & 6 \\
    0 & 0 & 0 & 8 & 6 & 2 \\
    0 & 0 & -12 & -6 & 12 & -6 \\
    0 & 0 & 6 & 2 & -6 & 4
    \end{bmatrix},
\end{eqnarray}
This matrix can be interpreted as adding a matrix $K_{bc}$ from the original $K_0$, where $K_{bc}$ is
\begin{eqnarray}
    K_{bc}=
        \begin{bmatrix}
        0  & -6  &  12  & -6  &  0  &  0  \\
        -6  &  0  &  6  & -2  &  0  &  0  \\
        12  &  6  &  0  &  0  &  0  &  0  \\
        -6  & -2  &  0  &  0  &  0  &  0  \\
        0  &  0  &  0  &  0  &  0  &  0  \\
        0  &  0  &  0  &  0  &  0  &  0  \\
        \end{bmatrix}.
\end{eqnarray}

\section{Evaluation of $\bra{\phi} K_{bc} \ket{\phi}$}
\label{app_sec:bc}
A simple example where $p=0, q=1$ is given by
\begin{eqnarray}
    \begin{bmatrix}
    0  & 1  &  0  & 0  &  0  &  0  \\
    1  &  0  &  0  & 0  &  0  &  0  \\
    0  &  0  &  0  &  0  &  0  &  0  \\
    0  & 0  &  0  &  0  &  0  &  0  \\
    0  &  0  &  0  &  0  &  0  &  0  \\
    0  &  0  &  0  &  0  &  0  &  0  \\
    \end{bmatrix},
\end{eqnarray}
The matrix representation are listed
\begin{eqnarray}
    K_{03} =
    \begin{bmatrix}
        0 & 0 & 0 & 1 \\
        0 & 0 & 0 & 0 \\
        0 & 0 & 0 & 0 \\
        1 & 0 & 0 & 0 \\
    \end{bmatrix},\\
    X(0) =
    \begin{bmatrix}
        0 & 1 & 0 & 0 \\
        1 & 0 & 0 & 0 \\
        0 & 0 & 0 & 1 \\
        0 & 0 & 1 & 0 \\
    \end{bmatrix},\\
    CNOT(0, 1) =
    \begin{bmatrix}
        1 & 0 & 0 & 0 \\
        0 & 0 & 0 & 1 \\
        0 & 0 & 1 & 0 \\
        0 & 1 & 0 & 0 \\
    \end{bmatrix},
\end{eqnarray}
The X(0) gate is applied to $K_{03}$ to transform it into $K_x$ as:
\begin{eqnarray}
    K_x = X(0)^T K_{03} X(0) =
    \begin{bmatrix}
        0 & 0 & 0 & 0 \\
        0 & 0 & 1 & 0 \\
        0 & 1 & 0 & 0 \\
        0 & 0 & 0 & 0 \\
    \end{bmatrix},
\end{eqnarray}
Then CNOT(0, 1) gate is applied to $K_x$, transforming it into $K_c$ as
\begin{eqnarray}
    K_c &=& CNOT(0,1)^T K_x CNOT(0, 1),\\
    &=&
    \begin{bmatrix}
        0 & 0 & 0 & 0 \\
        0 & 0 & 0 & 0 \\
        0 & 0 & 0 & 1 \\
        0 & 0 & 1 & 0 \\
    \end{bmatrix},\\ \nonumber
    &=& I_1 \otimes X = T^{T} K_{03} T. \nonumber
\end{eqnarray}

\section{Supplementary results of numerical experiments}
\label{app_sec:supp_plots}
\begin{figure}[hbt!]
\centering
\hspace{-1.4cm}
\begin{tikzpicture}[scale=1.]
    \begin{axis}[width=.6\textwidth, height=.3\textheight,
    ybar, bar width=0.5cm, enlarge x limits=0.15,
    legend style={at={(0.5,-0.15)},
    anchor=north,legend columns=-1},
    ytick={0,0.005,0.010,0.015},
    ymax=0.015,ymin=0, scaled y ticks=false,
    yticklabels={0,0.005,0.010,0.015},
    y label style={at={(-0.01,0.5)}},
    ylabel={Relative error},
    symbolic x coords={PBC, SSB, FFB, Cantilever}, xtick=data,
    xtick style={draw=none},
    ]
\addplot coordinates {(PBC,0.001210327) (SSB,0.008291119) (FFB,0.002510118)(Cantilever,0.014507971)};
\end{axis}
\end{tikzpicture}
\caption{Relative error between the target objective and predicted objective values across four boundary conditions.}
\label{fig:Relative_error}
\end{figure}
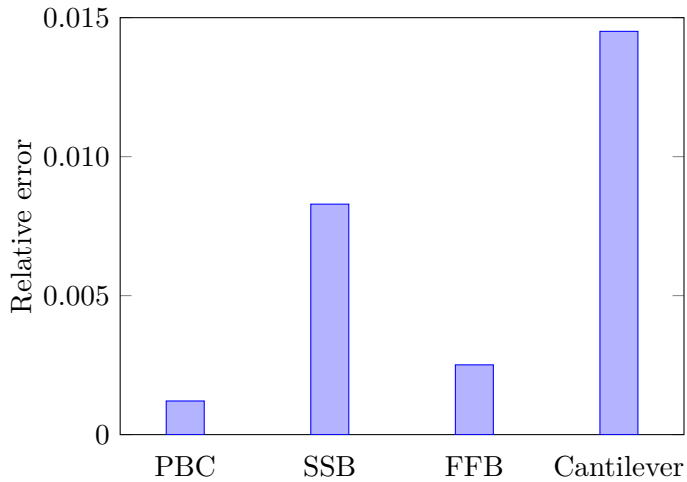

Fig.~\ref{fig:Relative_error} summarized the relative error between the predicted and target objective values for all the four cases.
\bibliography{ref}

\begin{thebibliography}{56}%
\makeatletter
\providecommand \@ifxundefined [1]{%
 \@ifx{#1\undefined}
}%
\providecommand \@ifnum [1]{%
 \ifnum #1\expandafter \@firstoftwo
 \else \expandafter \@secondoftwo
 \fi
}%
\providecommand \@ifx [1]{%
 \ifx #1\expandafter \@firstoftwo
 \else \expandafter \@secondoftwo
 \fi
}%
\providecommand \natexlab [1]{#1}%
\providecommand \enquote  [1]{``#1''}%
\providecommand \bibnamefont  [1]{#1}%
\providecommand \bibfnamefont [1]{#1}%
\providecommand \citenamefont [1]{#1}%
\providecommand \href@noop [0]{\@secondoftwo}%
\providecommand \href [0]{\begingroup \@sanitize@url \@href}%
\providecommand \@href[1]{\@@startlink{#1}\@@href}%
\providecommand \@@href[1]{\endgroup#1\@@endlink}%
\providecommand \@sanitize@url [0]{\catcode `\\12\catcode `\$12\catcode
  `\&12\catcode `\#12\catcode `\^12\catcode `\_12\catcode `\%12\relax}%
\providecommand \@@startlink[1]{}%
\providecommand \@@endlink[0]{}%
\providecommand \url  [0]{\begingroup\@sanitize@url \@url }%
\providecommand \@url [1]{\endgroup\@href {#1}{\urlprefix }}%
\providecommand \urlprefix  [0]{URL }%
\providecommand \Eprint [0]{\href }%
\providecommand \doibase [0]{https://doi.org/}%
\providecommand \selectlanguage [0]{\@gobble}%
\providecommand \bibinfo  [0]{\@secondoftwo}%
\providecommand \bibfield  [0]{\@secondoftwo}%
\providecommand \translation [1]{[#1]}%
\providecommand \BibitemOpen [0]{}%
\providecommand \bibitemStop [0]{}%
\providecommand \bibitemNoStop [0]{.\EOS\space}%
\providecommand \EOS [0]{\spacefactor3000\relax}%
\providecommand \BibitemShut  [1]{\csname bibitem#1\endcsname}%
\let\auto@bib@innerbib\@empty
\bibitem [{\citenamefont {Renardy}\ and\ \citenamefont
  {Rogers}(2006)}]{Renardy2006SSBM}%
  \BibitemOpen
  \bibfield  {author} {\bibinfo {author} {\bibfnamefont {M.}~\bibnamefont
  {Renardy}}\ and\ \bibinfo {author} {\bibfnamefont {R.~C.}\ \bibnamefont
  {Rogers}},\ }\href@noop {} {\emph {\bibinfo {title} {An introduction to
  partial differential equations}}},\ Vol.~\bibinfo {volume} {13}\ (\bibinfo
  {publisher} {Springer Science \& Business Media},\ \bibinfo {year}
  {2006})\BibitemShut {NoStop}%
\bibitem [{\citenamefont {Avalos}\ and\ \citenamefont
  {Geredeli}(2019)}]{Avalos2019MN}%
  \BibitemOpen
  \bibfield  {author} {\bibinfo {author} {\bibfnamefont {G.}~\bibnamefont
  {Avalos}}\ and\ \bibinfo {author} {\bibfnamefont {P.~G.}\ \bibnamefont
  {Geredeli}},\ }\href@noop {} {\bibfield  {journal} {\bibinfo  {journal}
  {Mathematische Nachrichten}\ }\textbf {\bibinfo {volume} {292}},\ \bibinfo
  {pages} {939} (\bibinfo {year} {2019})}\BibitemShut {NoStop}%
\bibitem [{\citenamefont {Rabczuk}\ \emph {et~al.}(2019)\citenamefont
  {Rabczuk}, \citenamefont {Ren},\ and\ \citenamefont
  {Zhuang}}]{Rabczuk2019CMC}%
  \BibitemOpen
  \bibfield  {author} {\bibinfo {author} {\bibfnamefont {T.}~\bibnamefont
  {Rabczuk}}, \bibinfo {author} {\bibfnamefont {H.}~\bibnamefont {Ren}},\ and\
  \bibinfo {author} {\bibfnamefont {X.}~\bibnamefont {Zhuang}},\ }\href@noop {}
  {\bibfield  {journal} {\bibinfo  {journal} {Computers, Materials \& Continua
  59 (2019), Nr. 1}\ ,\ \bibinfo {pages} {31}} (\bibinfo {year}
  {2019})}\BibitemShut {NoStop}%
\bibitem [{\citenamefont {Brion}\ \emph {et~al.}(2023)\citenamefont {Brion},
  \citenamefont {Fossat}, \citenamefont {Ichchou}, \citenamefont {Bareille},
  \citenamefont {Zine},\ and\ \citenamefont {Droz}}]{Brion2023CS}%
  \BibitemOpen
  \bibfield  {author} {\bibinfo {author} {\bibfnamefont {T.}~\bibnamefont
  {Brion}}, \bibinfo {author} {\bibfnamefont {P.}~\bibnamefont {Fossat}},
  \bibinfo {author} {\bibfnamefont {M.}~\bibnamefont {Ichchou}}, \bibinfo
  {author} {\bibfnamefont {O.}~\bibnamefont {Bareille}}, \bibinfo {author}
  {\bibfnamefont {A.-M.}\ \bibnamefont {Zine}},\ and\ \bibinfo {author}
  {\bibfnamefont {C.}~\bibnamefont {Droz}},\ }\href@noop {} {\bibfield
  {journal} {\bibinfo  {journal} {Composite Structures}\ }\textbf {\bibinfo
  {volume} {304}},\ \bibinfo {pages} {116297} (\bibinfo {year}
  {2023})}\BibitemShut {NoStop}%
\bibitem [{\citenamefont {Klawonn}\ and\ \citenamefont
  {Rheinbach}(2010)}]{Klawonn2010ZAMM}%
  \BibitemOpen
  \bibfield  {author} {\bibinfo {author} {\bibfnamefont {A.}~\bibnamefont
  {Klawonn}}\ and\ \bibinfo {author} {\bibfnamefont {O.}~\bibnamefont
  {Rheinbach}},\ }\href@noop {} {\bibfield  {journal} {\bibinfo  {journal}
  {ZAMM-Journal of Applied Mathematics and Mechanics/Zeitschrift f{\"u}r
  Angewandte Mathematik und Mechanik: Applied Mathematics and Mechanics}\
  }\textbf {\bibinfo {volume} {90}},\ \bibinfo {pages} {5} (\bibinfo {year}
  {2010})}\BibitemShut {NoStop}%
\bibitem [{\citenamefont {Klawonn}\ \emph {et~al.}(2015)\citenamefont
  {Klawonn}, \citenamefont {Lanser},\ and\ \citenamefont
  {Rheinbach}}]{Klawonn2015SIAM}%
  \BibitemOpen
  \bibfield  {author} {\bibinfo {author} {\bibfnamefont {A.}~\bibnamefont
  {Klawonn}}, \bibinfo {author} {\bibfnamefont {M.}~\bibnamefont {Lanser}},\
  and\ \bibinfo {author} {\bibfnamefont {O.}~\bibnamefont {Rheinbach}},\
  }\href@noop {} {\bibfield  {journal} {\bibinfo  {journal} {SIAM Journal on
  Scientific Computing}\ }\textbf {\bibinfo {volume} {37}},\ \bibinfo {pages}
  {C667} (\bibinfo {year} {2015})}\BibitemShut {NoStop}%
\bibitem [{\citenamefont {Toivanen}\ \emph {et~al.}(2018)\citenamefont
  {Toivanen}, \citenamefont {Avery},\ and\ \citenamefont
  {Farhat}}]{Toivanen2018IJNME}%
  \BibitemOpen
  \bibfield  {author} {\bibinfo {author} {\bibfnamefont {J.}~\bibnamefont
  {Toivanen}}, \bibinfo {author} {\bibfnamefont {P.}~\bibnamefont {Avery}},\
  and\ \bibinfo {author} {\bibfnamefont {C.}~\bibnamefont {Farhat}},\
  }\href@noop {} {\bibfield  {journal} {\bibinfo  {journal} {International
  Journal for Numerical Methods in Engineering}\ }\textbf {\bibinfo {volume}
  {116}},\ \bibinfo {pages} {661} (\bibinfo {year} {2018})}\BibitemShut
  {NoStop}%
\bibitem [{\citenamefont {Fujita}\ \emph {et~al.}(2021)\citenamefont {Fujita},
  \citenamefont {Koyama}, \citenamefont {Minami}, \citenamefont {Inoue},
  \citenamefont {Nishizawa}, \citenamefont {Tsuji}, \citenamefont {Nishiki},
  \citenamefont {Ichimura}, \citenamefont {Hori},\ and\ \citenamefont
  {Maddegedara}}]{Fujita2021JOCS}%
  \BibitemOpen
  \bibfield  {author} {\bibinfo {author} {\bibfnamefont {K.}~\bibnamefont
  {Fujita}}, \bibinfo {author} {\bibfnamefont {K.}~\bibnamefont {Koyama}},
  \bibinfo {author} {\bibfnamefont {K.}~\bibnamefont {Minami}}, \bibinfo
  {author} {\bibfnamefont {H.}~\bibnamefont {Inoue}}, \bibinfo {author}
  {\bibfnamefont {S.}~\bibnamefont {Nishizawa}}, \bibinfo {author}
  {\bibfnamefont {M.}~\bibnamefont {Tsuji}}, \bibinfo {author} {\bibfnamefont
  {T.}~\bibnamefont {Nishiki}}, \bibinfo {author} {\bibfnamefont
  {T.}~\bibnamefont {Ichimura}}, \bibinfo {author} {\bibfnamefont
  {M.}~\bibnamefont {Hori}},\ and\ \bibinfo {author} {\bibfnamefont
  {L.}~\bibnamefont {Maddegedara}},\ }\href@noop {} {\bibfield  {journal}
  {\bibinfo  {journal} {Journal of Computational Science}\ }\textbf {\bibinfo
  {volume} {49}},\ \bibinfo {pages} {101277} (\bibinfo {year}
  {2021})}\BibitemShut {NoStop}%
\bibitem [{\citenamefont {of~Sciences~Engineering}\ \emph
  {et~al.}(2019)\citenamefont {of~Sciences~Engineering}, \citenamefont
  {Medicine} \emph {et~al.}}]{Horowitz2019}%
  \BibitemOpen
  \bibfield  {author} {\bibinfo {author} {\bibfnamefont {N.~A.}\ \bibnamefont
  {of~Sciences~Engineering}}, \bibinfo {author} {\bibnamefont {Medicine}},
  \emph {et~al.},\ }\href@noop {} {\emph {\bibinfo {title} {Quantum computing:
  progress and prospects}}}\ (\bibinfo  {publisher} {National Academies
  Press},\ \bibinfo {year} {2019})\BibitemShut {NoStop}%
\bibitem [{\citenamefont {Haug}\ \emph {et~al.}(2021)\citenamefont {Haug},
  \citenamefont {Bharti},\ and\ \citenamefont {Kim}}]{Haug2021PRX}%
  \BibitemOpen
  \bibfield  {author} {\bibinfo {author} {\bibfnamefont {T.}~\bibnamefont
  {Haug}}, \bibinfo {author} {\bibfnamefont {K.}~\bibnamefont {Bharti}},\ and\
  \bibinfo {author} {\bibfnamefont {M.}~\bibnamefont {Kim}},\ }\href@noop {}
  {\bibfield  {journal} {\bibinfo  {journal} {PRX Quantum}\ }\textbf {\bibinfo
  {volume} {2}},\ \bibinfo {pages} {040309} (\bibinfo {year}
  {2021})}\BibitemShut {NoStop}%
\bibitem [{\citenamefont {Shor}(1994)}]{Shor1994}%
  \BibitemOpen
  \bibfield  {author} {\bibinfo {author} {\bibfnamefont {P.~W.}\ \bibnamefont
  {Shor}},\ }in\ \href@noop {} {\emph {\bibinfo {booktitle} {Proceedings 35th
  annual symposium on foundations of computer science}}}\ (\bibinfo
  {organization} {Ieee},\ \bibinfo {year} {1994})\ pp.\ \bibinfo {pages}
  {124--134}\BibitemShut {NoStop}%
\bibitem [{\citenamefont {Harrow}\ \emph {et~al.}(2009)\citenamefont {Harrow},
  \citenamefont {Hassidim},\ and\ \citenamefont {Lloyd}}]{Harrow2009PRL}%
  \BibitemOpen
  \bibfield  {author} {\bibinfo {author} {\bibfnamefont {A.~W.}\ \bibnamefont
  {Harrow}}, \bibinfo {author} {\bibfnamefont {A.}~\bibnamefont {Hassidim}},\
  and\ \bibinfo {author} {\bibfnamefont {S.}~\bibnamefont {Lloyd}},\
  }\href@noop {} {\bibfield  {journal} {\bibinfo  {journal} {Physical review
  letters}\ }\textbf {\bibinfo {volume} {103}},\ \bibinfo {pages} {150502}
  (\bibinfo {year} {2009})}\BibitemShut {NoStop}%
\bibitem [{\citenamefont {Cao}\ \emph {et~al.}(2012)\citenamefont {Cao},
  \citenamefont {Daskin}, \citenamefont {Frankel},\ and\ \citenamefont
  {Kais}}]{Cao2012MP}%
  \BibitemOpen
  \bibfield  {author} {\bibinfo {author} {\bibfnamefont {Y.}~\bibnamefont
  {Cao}}, \bibinfo {author} {\bibfnamefont {A.}~\bibnamefont {Daskin}},
  \bibinfo {author} {\bibfnamefont {S.}~\bibnamefont {Frankel}},\ and\ \bibinfo
  {author} {\bibfnamefont {S.}~\bibnamefont {Kais}},\ }\href@noop {} {\bibfield
   {journal} {\bibinfo  {journal} {Molecular Physics}\ }\textbf {\bibinfo
  {volume} {110}},\ \bibinfo {pages} {1675} (\bibinfo {year}
  {2012})}\BibitemShut {NoStop}%
\bibitem [{\citenamefont {Clader}\ \emph {et~al.}(2013)\citenamefont {Clader},
  \citenamefont {Jacobs},\ and\ \citenamefont {Sprouse}}]{Clader2013PRL}%
  \BibitemOpen
  \bibfield  {author} {\bibinfo {author} {\bibfnamefont {B.~D.}\ \bibnamefont
  {Clader}}, \bibinfo {author} {\bibfnamefont {B.~C.}\ \bibnamefont {Jacobs}},\
  and\ \bibinfo {author} {\bibfnamefont {C.~R.}\ \bibnamefont {Sprouse}},\
  }\href@noop {} {\bibfield  {journal} {\bibinfo  {journal} {Physical review
  letters}\ }\textbf {\bibinfo {volume} {110}},\ \bibinfo {pages} {250504}
  (\bibinfo {year} {2013})}\BibitemShut {NoStop}%
\bibitem [{\citenamefont {Zhang}\ \emph {et~al.}(2021)\citenamefont {Zhang},
  \citenamefont {Feng},\ and\ \citenamefont {Zhang}}]{Zhang2021IEEE}%
  \BibitemOpen
  \bibfield  {author} {\bibinfo {author} {\bibfnamefont {J.}~\bibnamefont
  {Zhang}}, \bibinfo {author} {\bibfnamefont {F.}~\bibnamefont {Feng}},\ and\
  \bibinfo {author} {\bibfnamefont {Q.}~\bibnamefont {Zhang}},\ }in\ \href@noop
  {} {\emph {\bibinfo {booktitle} {2021 IEEE MTT-S International Microwave
  Symposium (IMS)}}}\ (\bibinfo {organization} {IEEE},\ \bibinfo {year}
  {2021})\ pp.\ \bibinfo {pages} {120--123}\BibitemShut {NoStop}%
\bibitem [{\citenamefont {Cao}\ \emph {et~al.}(2013)\citenamefont {Cao},
  \citenamefont {Papageorgiou}, \citenamefont {Petras}, \citenamefont {Traub},\
  and\ \citenamefont {Kais}}]{Cao2013NJP}%
  \BibitemOpen
  \bibfield  {author} {\bibinfo {author} {\bibfnamefont {Y.}~\bibnamefont
  {Cao}}, \bibinfo {author} {\bibfnamefont {A.}~\bibnamefont {Papageorgiou}},
  \bibinfo {author} {\bibfnamefont {I.}~\bibnamefont {Petras}}, \bibinfo
  {author} {\bibfnamefont {J.}~\bibnamefont {Traub}},\ and\ \bibinfo {author}
  {\bibfnamefont {S.}~\bibnamefont {Kais}},\ }\href@noop {} {\bibfield
  {journal} {\bibinfo  {journal} {New Journal of Physics}\ }\textbf {\bibinfo
  {volume} {15}},\ \bibinfo {pages} {013021} (\bibinfo {year}
  {2013})}\BibitemShut {NoStop}%
\bibitem [{\citenamefont {Sato}\ \emph {et~al.}(2021)\citenamefont {Sato},
  \citenamefont {Kondo}, \citenamefont {Koide}, \citenamefont {Takamatsu},\
  and\ \citenamefont {Imoto}}]{Sato2021PRA}%
  \BibitemOpen
  \bibfield  {author} {\bibinfo {author} {\bibfnamefont {Y.}~\bibnamefont
  {Sato}}, \bibinfo {author} {\bibfnamefont {R.}~\bibnamefont {Kondo}},
  \bibinfo {author} {\bibfnamefont {S.}~\bibnamefont {Koide}}, \bibinfo
  {author} {\bibfnamefont {H.}~\bibnamefont {Takamatsu}},\ and\ \bibinfo
  {author} {\bibfnamefont {N.}~\bibnamefont {Imoto}},\ }\href@noop {}
  {\bibfield  {journal} {\bibinfo  {journal} {Physical Review A}\ }\textbf
  {\bibinfo {volume} {104}},\ \bibinfo {pages} {052409} (\bibinfo {year}
  {2021})}\BibitemShut {NoStop}%
\bibitem [{\citenamefont {Liu}\ \emph {et~al.}(2021)\citenamefont {Liu},
  \citenamefont {Wu}, \citenamefont {Wan}, \citenamefont {Pan}, \citenamefont
  {Qin}, \citenamefont {Gao},\ and\ \citenamefont {Wen}}]{Liu2021PRA}%
  \BibitemOpen
  \bibfield  {author} {\bibinfo {author} {\bibfnamefont {H.-L.}\ \bibnamefont
  {Liu}}, \bibinfo {author} {\bibfnamefont {Y.-S.}\ \bibnamefont {Wu}},
  \bibinfo {author} {\bibfnamefont {L.-C.}\ \bibnamefont {Wan}}, \bibinfo
  {author} {\bibfnamefont {S.-J.}\ \bibnamefont {Pan}}, \bibinfo {author}
  {\bibfnamefont {S.-J.}\ \bibnamefont {Qin}}, \bibinfo {author} {\bibfnamefont
  {F.}~\bibnamefont {Gao}},\ and\ \bibinfo {author} {\bibfnamefont {Q.-Y.}\
  \bibnamefont {Wen}},\ }\href@noop {} {\bibfield  {journal} {\bibinfo
  {journal} {Physical Review A}\ }\textbf {\bibinfo {volume} {104}},\ \bibinfo
  {pages} {022418} (\bibinfo {year} {2021})}\BibitemShut {NoStop}%
\bibitem [{\citenamefont {Bravo-Prieto}\ \emph {et~al.}(2019)\citenamefont
  {Bravo-Prieto}, \citenamefont {LaRose}, \citenamefont {Cerezo}, \citenamefont
  {Subasi}, \citenamefont {Cincio},\ and\ \citenamefont
  {Coles}}]{Bravo2019arXiv}%
  \BibitemOpen
  \bibfield  {author} {\bibinfo {author} {\bibfnamefont {C.}~\bibnamefont
  {Bravo-Prieto}}, \bibinfo {author} {\bibfnamefont {R.}~\bibnamefont
  {LaRose}}, \bibinfo {author} {\bibfnamefont {M.}~\bibnamefont {Cerezo}},
  \bibinfo {author} {\bibfnamefont {Y.}~\bibnamefont {Subasi}}, \bibinfo
  {author} {\bibfnamefont {L.}~\bibnamefont {Cincio}},\ and\ \bibinfo {author}
  {\bibfnamefont {P.~J.}\ \bibnamefont {Coles}},\ }\href@noop {} {\bibfield
  {journal} {\bibinfo  {journal} {arXiv preprint arXiv:1909.05820}\ } (\bibinfo
  {year} {2019})}\BibitemShut {NoStop}%
\bibitem [{\citenamefont {Preskill}(2018)}]{Preskill2018Quantum}%
  \BibitemOpen
  \bibfield  {author} {\bibinfo {author} {\bibfnamefont {J.~J.~Q.}\
  \bibnamefont {Preskill}},\ }\href@noop {} {\bibfield  {journal} {\bibinfo
  {journal} {Quantum}\ }\textbf {\bibinfo {volume} {2}},\ \bibinfo {pages} {79}
  (\bibinfo {year} {2018})}\BibitemShut {NoStop}%
\bibitem [{\citenamefont {Cerezo}\ \emph {et~al.}(2021)\citenamefont {Cerezo},
  \citenamefont {Arrasmith}, \citenamefont {Babbush}, \citenamefont {Benjamin},
  \citenamefont {Endo}, \citenamefont {Fujii}, \citenamefont {McClean},
  \citenamefont {Mitarai}, \citenamefont {Yuan}, \citenamefont {Cincio} \emph
  {et~al.}}]{Cerezo2021NRP}%
  \BibitemOpen
  \bibfield  {author} {\bibinfo {author} {\bibfnamefont {M.}~\bibnamefont
  {Cerezo}}, \bibinfo {author} {\bibfnamefont {A.}~\bibnamefont {Arrasmith}},
  \bibinfo {author} {\bibfnamefont {R.}~\bibnamefont {Babbush}}, \bibinfo
  {author} {\bibfnamefont {S.~C.}\ \bibnamefont {Benjamin}}, \bibinfo {author}
  {\bibfnamefont {S.}~\bibnamefont {Endo}}, \bibinfo {author} {\bibfnamefont
  {K.}~\bibnamefont {Fujii}}, \bibinfo {author} {\bibfnamefont {J.~R.}\
  \bibnamefont {McClean}}, \bibinfo {author} {\bibfnamefont {K.}~\bibnamefont
  {Mitarai}}, \bibinfo {author} {\bibfnamefont {X.}~\bibnamefont {Yuan}},
  \bibinfo {author} {\bibfnamefont {L.}~\bibnamefont {Cincio}}, \emph
  {et~al.},\ }\href@noop {} {\bibfield  {journal} {\bibinfo  {journal} {Nature
  Reviews Physics}\ }\textbf {\bibinfo {volume} {3}},\ \bibinfo {pages} {625}
  (\bibinfo {year} {2021})}\BibitemShut {NoStop}%
\bibitem [{\citenamefont {Enomoto}\ \emph {et~al.}(2022)\citenamefont
  {Enomoto}, \citenamefont {Anai}, \citenamefont {Udagawa},\ and\ \citenamefont
  {Takeda}}]{Enomoto2022PRR}%
  \BibitemOpen
  \bibfield  {author} {\bibinfo {author} {\bibfnamefont {Y.}~\bibnamefont
  {Enomoto}}, \bibinfo {author} {\bibfnamefont {K.}~\bibnamefont {Anai}},
  \bibinfo {author} {\bibfnamefont {K.}~\bibnamefont {Udagawa}},\ and\ \bibinfo
  {author} {\bibfnamefont {S.}~\bibnamefont {Takeda}},\ }\href@noop {}
  {\bibfield  {journal} {\bibinfo  {journal} {PHYSICAL REVIEW RESEARCH}\ }
  (\bibinfo {year} {2022})}\BibitemShut {NoStop}%
\bibitem [{\citenamefont {Holmes}\ \emph {et~al.}(2022)\citenamefont {Holmes},
  \citenamefont {Sharma}, \citenamefont {Cerezo},\ and\ \citenamefont
  {Coles}}]{Holmes2022PRX}%
  \BibitemOpen
  \bibfield  {author} {\bibinfo {author} {\bibfnamefont {Z.}~\bibnamefont
  {Holmes}}, \bibinfo {author} {\bibfnamefont {K.}~\bibnamefont {Sharma}},
  \bibinfo {author} {\bibfnamefont {M.}~\bibnamefont {Cerezo}},\ and\ \bibinfo
  {author} {\bibfnamefont {P.~J.}\ \bibnamefont {Coles}},\ }\href@noop {}
  {\bibfield  {journal} {\bibinfo  {journal} {PRX Quantum}\ }\textbf {\bibinfo
  {volume} {3}},\ \bibinfo {pages} {010313} (\bibinfo {year}
  {2022})}\BibitemShut {NoStop}%
\bibitem [{\citenamefont {Fedorov}\ \emph {et~al.}(2022)\citenamefont
  {Fedorov}, \citenamefont {Peng}, \citenamefont {Govind},\ and\ \citenamefont
  {Alexeev}}]{Fedorov2022MT}%
  \BibitemOpen
  \bibfield  {author} {\bibinfo {author} {\bibfnamefont {D.~A.}\ \bibnamefont
  {Fedorov}}, \bibinfo {author} {\bibfnamefont {B.}~\bibnamefont {Peng}},
  \bibinfo {author} {\bibfnamefont {N.}~\bibnamefont {Govind}},\ and\ \bibinfo
  {author} {\bibfnamefont {Y.}~\bibnamefont {Alexeev}},\ }\href@noop {}
  {\bibfield  {journal} {\bibinfo  {journal} {Materials Theory}\ }\textbf
  {\bibinfo {volume} {6}},\ \bibinfo {pages} {1} (\bibinfo {year}
  {2022})}\BibitemShut {NoStop}%
\bibitem [{\citenamefont {Cervera-Lierta}\ \emph {et~al.}(2021)\citenamefont
  {Cervera-Lierta}, \citenamefont {Kottmann},\ and\ \citenamefont
  {Aspuru-Guzik}}]{Cervera2021PRX}%
  \BibitemOpen
  \bibfield  {author} {\bibinfo {author} {\bibfnamefont {A.}~\bibnamefont
  {Cervera-Lierta}}, \bibinfo {author} {\bibfnamefont {J.~S.}\ \bibnamefont
  {Kottmann}},\ and\ \bibinfo {author} {\bibfnamefont {A.}~\bibnamefont
  {Aspuru-Guzik}},\ }\href@noop {} {\bibfield  {journal} {\bibinfo  {journal}
  {PRX Quantum}\ }\textbf {\bibinfo {volume} {2}},\ \bibinfo {pages} {020329}
  (\bibinfo {year} {2021})}\BibitemShut {NoStop}%
\bibitem [{\citenamefont {Magann}\ \emph {et~al.}(2021)\citenamefont {Magann},
  \citenamefont {Arenz}, \citenamefont {Grace}, \citenamefont {Ho},
  \citenamefont {Kosut}, \citenamefont {McClean}, \citenamefont {Rabitz},\ and\
  \citenamefont {Sarovar}}]{Magann2021PRX}%
  \BibitemOpen
  \bibfield  {author} {\bibinfo {author} {\bibfnamefont {A.~B.}\ \bibnamefont
  {Magann}}, \bibinfo {author} {\bibfnamefont {C.}~\bibnamefont {Arenz}},
  \bibinfo {author} {\bibfnamefont {M.~D.}\ \bibnamefont {Grace}}, \bibinfo
  {author} {\bibfnamefont {T.-S.}\ \bibnamefont {Ho}}, \bibinfo {author}
  {\bibfnamefont {R.~L.}\ \bibnamefont {Kosut}}, \bibinfo {author}
  {\bibfnamefont {J.~R.}\ \bibnamefont {McClean}}, \bibinfo {author}
  {\bibfnamefont {H.~A.}\ \bibnamefont {Rabitz}},\ and\ \bibinfo {author}
  {\bibfnamefont {M.}~\bibnamefont {Sarovar}},\ }\href@noop {} {\bibfield
  {journal} {\bibinfo  {journal} {PRX Quantum}\ }\textbf {\bibinfo {volume}
  {2}},\ \bibinfo {pages} {010101} (\bibinfo {year} {2021})}\BibitemShut
  {NoStop}%
\bibitem [{\citenamefont {Peruzzo}\ \emph {et~al.}(2014)\citenamefont
  {Peruzzo}, \citenamefont {McClean}, \citenamefont {Shadbolt}, \citenamefont
  {Yung}, \citenamefont {Zhou}, \citenamefont {Love}, \citenamefont
  {Aspuru-Guzik},\ and\ \citenamefont {O’brien}}]{Peruzzo2014Nature}%
  \BibitemOpen
  \bibfield  {author} {\bibinfo {author} {\bibfnamefont {A.}~\bibnamefont
  {Peruzzo}}, \bibinfo {author} {\bibfnamefont {J.}~\bibnamefont {McClean}},
  \bibinfo {author} {\bibfnamefont {P.}~\bibnamefont {Shadbolt}}, \bibinfo
  {author} {\bibfnamefont {M.-H.}\ \bibnamefont {Yung}}, \bibinfo {author}
  {\bibfnamefont {X.-Q.}\ \bibnamefont {Zhou}}, \bibinfo {author}
  {\bibfnamefont {P.~J.}\ \bibnamefont {Love}}, \bibinfo {author}
  {\bibfnamefont {A.}~\bibnamefont {Aspuru-Guzik}},\ and\ \bibinfo {author}
  {\bibfnamefont {J.~L.}\ \bibnamefont {O’brien}},\ }\href@noop {} {\bibfield
   {journal} {\bibinfo  {journal} {Nature communications}\ }\textbf {\bibinfo
  {volume} {5}},\ \bibinfo {pages} {4213} (\bibinfo {year} {2014})}\BibitemShut
  {NoStop}%
\bibitem [{\citenamefont {Tilly}\ \emph {et~al.}(2022)\citenamefont {Tilly},
  \citenamefont {Chen}, \citenamefont {Cao}, \citenamefont {Picozzi},
  \citenamefont {Setia}, \citenamefont {Li}, \citenamefont {Grant},
  \citenamefont {Wossnig}, \citenamefont {Rungger}, \citenamefont {Booth} \emph
  {et~al.}}]{Tilly2022PR}%
  \BibitemOpen
  \bibfield  {author} {\bibinfo {author} {\bibfnamefont {J.}~\bibnamefont
  {Tilly}}, \bibinfo {author} {\bibfnamefont {H.}~\bibnamefont {Chen}},
  \bibinfo {author} {\bibfnamefont {S.}~\bibnamefont {Cao}}, \bibinfo {author}
  {\bibfnamefont {D.}~\bibnamefont {Picozzi}}, \bibinfo {author} {\bibfnamefont
  {K.}~\bibnamefont {Setia}}, \bibinfo {author} {\bibfnamefont
  {Y.}~\bibnamefont {Li}}, \bibinfo {author} {\bibfnamefont {E.}~\bibnamefont
  {Grant}}, \bibinfo {author} {\bibfnamefont {L.}~\bibnamefont {Wossnig}},
  \bibinfo {author} {\bibfnamefont {I.}~\bibnamefont {Rungger}}, \bibinfo
  {author} {\bibfnamefont {G.~H.}\ \bibnamefont {Booth}}, \emph {et~al.},\
  }\href@noop {} {\bibfield  {journal} {\bibinfo  {journal} {Physics Reports}\
  }\textbf {\bibinfo {volume} {986}},\ \bibinfo {pages} {1} (\bibinfo {year}
  {2022})}\BibitemShut {NoStop}%
\bibitem [{\citenamefont {Farhi}\ \emph {et~al.}(2014)\citenamefont {Farhi},
  \citenamefont {Goldstone},\ and\ \citenamefont {Gutmann}}]{Farhi2014arXiv}%
  \BibitemOpen
  \bibfield  {author} {\bibinfo {author} {\bibfnamefont {E.}~\bibnamefont
  {Farhi}}, \bibinfo {author} {\bibfnamefont {J.}~\bibnamefont {Goldstone}},\
  and\ \bibinfo {author} {\bibfnamefont {S.}~\bibnamefont {Gutmann}},\
  }\href@noop {} {\bibfield  {journal} {\bibinfo  {journal} {arXiv preprint
  arXiv:1411.4028}\ } (\bibinfo {year} {2014})}\BibitemShut {NoStop}%
\bibitem [{\citenamefont {Hadfield}\ \emph {et~al.}(2019)\citenamefont
  {Hadfield}, \citenamefont {Wang}, \citenamefont {O’gorman}, \citenamefont
  {Rieffel}, \citenamefont {Venturelli},\ and\ \citenamefont
  {Biswas}}]{Hadfield2019Algorithms}%
  \BibitemOpen
  \bibfield  {author} {\bibinfo {author} {\bibfnamefont {S.}~\bibnamefont
  {Hadfield}}, \bibinfo {author} {\bibfnamefont {Z.}~\bibnamefont {Wang}},
  \bibinfo {author} {\bibfnamefont {B.}~\bibnamefont {O’gorman}}, \bibinfo
  {author} {\bibfnamefont {E.~G.}\ \bibnamefont {Rieffel}}, \bibinfo {author}
  {\bibfnamefont {D.}~\bibnamefont {Venturelli}},\ and\ \bibinfo {author}
  {\bibfnamefont {R.}~\bibnamefont {Biswas}},\ }\href@noop {} {\bibfield
  {journal} {\bibinfo  {journal} {Algorithms}\ }\textbf {\bibinfo {volume}
  {12}},\ \bibinfo {pages} {34} (\bibinfo {year} {2019})}\BibitemShut {NoStop}%
\bibitem [{\citenamefont {Alexeev}\ \emph {et~al.}(2021)\citenamefont
  {Alexeev}, \citenamefont {Bacon}, \citenamefont {Brown}, \citenamefont
  {Calderbank}, \citenamefont {Carr}, \citenamefont {Chong}, \citenamefont
  {DeMarco}, \citenamefont {Englund}, \citenamefont {Farhi}, \citenamefont
  {Fefferman} \emph {et~al.}}]{Alexeev2021PRX}%
  \BibitemOpen
  \bibfield  {author} {\bibinfo {author} {\bibfnamefont {Y.}~\bibnamefont
  {Alexeev}}, \bibinfo {author} {\bibfnamefont {D.}~\bibnamefont {Bacon}},
  \bibinfo {author} {\bibfnamefont {K.~R.}\ \bibnamefont {Brown}}, \bibinfo
  {author} {\bibfnamefont {R.}~\bibnamefont {Calderbank}}, \bibinfo {author}
  {\bibfnamefont {L.~D.}\ \bibnamefont {Carr}}, \bibinfo {author}
  {\bibfnamefont {F.~T.}\ \bibnamefont {Chong}}, \bibinfo {author}
  {\bibfnamefont {B.}~\bibnamefont {DeMarco}}, \bibinfo {author} {\bibfnamefont
  {D.}~\bibnamefont {Englund}}, \bibinfo {author} {\bibfnamefont
  {E.}~\bibnamefont {Farhi}}, \bibinfo {author} {\bibfnamefont
  {B.}~\bibnamefont {Fefferman}}, \emph {et~al.},\ }\href@noop {} {\bibfield
  {journal} {\bibinfo  {journal} {PRX Quantum}\ }\textbf {\bibinfo {volume}
  {2}},\ \bibinfo {pages} {017001} (\bibinfo {year} {2021})}\BibitemShut
  {NoStop}%
\bibitem [{\citenamefont {Zhou}\ \emph {et~al.}(2020)\citenamefont {Zhou},
  \citenamefont {Wang}, \citenamefont {Choi}, \citenamefont {Pichler},\ and\
  \citenamefont {Lukin}}]{Zhou2020PRX}%
  \BibitemOpen
  \bibfield  {author} {\bibinfo {author} {\bibfnamefont {L.}~\bibnamefont
  {Zhou}}, \bibinfo {author} {\bibfnamefont {S.-T.}\ \bibnamefont {Wang}},
  \bibinfo {author} {\bibfnamefont {S.}~\bibnamefont {Choi}}, \bibinfo {author}
  {\bibfnamefont {H.}~\bibnamefont {Pichler}},\ and\ \bibinfo {author}
  {\bibfnamefont {M.~D.}\ \bibnamefont {Lukin}},\ }\href@noop {} {\bibfield
  {journal} {\bibinfo  {journal} {Physical Review X}\ }\textbf {\bibinfo
  {volume} {10}},\ \bibinfo {pages} {021067} (\bibinfo {year}
  {2020})}\BibitemShut {NoStop}%
\bibitem [{\citenamefont {Matos}\ \emph {et~al.}(2021)\citenamefont {Matos},
  \citenamefont {Johri},\ and\ \citenamefont {Papi{\'c}}}]{Matos2021PRX}%
  \BibitemOpen
  \bibfield  {author} {\bibinfo {author} {\bibfnamefont {G.}~\bibnamefont
  {Matos}}, \bibinfo {author} {\bibfnamefont {S.}~\bibnamefont {Johri}},\ and\
  \bibinfo {author} {\bibfnamefont {Z.}~\bibnamefont {Papi{\'c}}},\ }\href@noop
  {} {\bibfield  {journal} {\bibinfo  {journal} {PRX Quantum}\ }\textbf
  {\bibinfo {volume} {2}},\ \bibinfo {pages} {010309} (\bibinfo {year}
  {2021})}\BibitemShut {NoStop}%
\bibitem [{\citenamefont {Xu}\ \emph {et~al.}(2021)\citenamefont {Xu},
  \citenamefont {Sun}, \citenamefont {Endo}, \citenamefont {Li}, \citenamefont
  {Benjamin},\ and\ \citenamefont {Yuan}}]{Xu2021SB}%
  \BibitemOpen
  \bibfield  {author} {\bibinfo {author} {\bibfnamefont {X.}~\bibnamefont
  {Xu}}, \bibinfo {author} {\bibfnamefont {J.}~\bibnamefont {Sun}}, \bibinfo
  {author} {\bibfnamefont {S.}~\bibnamefont {Endo}}, \bibinfo {author}
  {\bibfnamefont {Y.}~\bibnamefont {Li}}, \bibinfo {author} {\bibfnamefont
  {S.~C.}\ \bibnamefont {Benjamin}},\ and\ \bibinfo {author} {\bibfnamefont
  {X.}~\bibnamefont {Yuan}},\ }\href@noop {} {\bibfield  {journal} {\bibinfo
  {journal} {Science Bulletin}\ }\textbf {\bibinfo {volume} {66}},\ \bibinfo
  {pages} {2181} (\bibinfo {year} {2021})}\BibitemShut {NoStop}%
\bibitem [{\citenamefont {Huang}\ \emph {et~al.}(2021)\citenamefont {Huang},
  \citenamefont {Bharti},\ and\ \citenamefont {Rebentrost}}]{Huang2021NJP}%
  \BibitemOpen
  \bibfield  {author} {\bibinfo {author} {\bibfnamefont {H.-Y.}\ \bibnamefont
  {Huang}}, \bibinfo {author} {\bibfnamefont {K.}~\bibnamefont {Bharti}},\ and\
  \bibinfo {author} {\bibfnamefont {P.}~\bibnamefont {Rebentrost}},\
  }\href@noop {} {\bibfield  {journal} {\bibinfo  {journal} {New Journal of
  Physics}\ }\textbf {\bibinfo {volume} {23}},\ \bibinfo {pages} {113021}
  (\bibinfo {year} {2021})}\BibitemShut {NoStop}%
\bibitem [{\citenamefont {Ewe}\ \emph {et~al.}(2022)\citenamefont {Ewe},
  \citenamefont {Koh}, \citenamefont {Goh}, \citenamefont {Chu},\ and\
  \citenamefont {Png}}]{Ewe2022IEEE}%
  \BibitemOpen
  \bibfield  {author} {\bibinfo {author} {\bibfnamefont {W.-B.}\ \bibnamefont
  {Ewe}}, \bibinfo {author} {\bibfnamefont {D.~E.}\ \bibnamefont {Koh}},
  \bibinfo {author} {\bibfnamefont {S.~T.}\ \bibnamefont {Goh}}, \bibinfo
  {author} {\bibfnamefont {H.-S.}\ \bibnamefont {Chu}},\ and\ \bibinfo {author}
  {\bibfnamefont {C.~E.}\ \bibnamefont {Png}},\ }\href@noop {} {\bibfield
  {journal} {\bibinfo  {journal} {IEEE Transactions on Microwave Theory and
  Techniques}\ }\textbf {\bibinfo {volume} {70}},\ \bibinfo {pages} {2517}
  (\bibinfo {year} {2022})}\BibitemShut {NoStop}%
\bibitem [{\citenamefont {Manteuffel}(1980)}]{Manteuffel1980MOC}%
  \BibitemOpen
  \bibfield  {author} {\bibinfo {author} {\bibfnamefont {T.~A.}\ \bibnamefont
  {Manteuffel}},\ }\href@noop {} {\bibfield  {journal} {\bibinfo  {journal}
  {Mathematics of computation}\ }\textbf {\bibinfo {volume} {34}},\ \bibinfo
  {pages} {473} (\bibinfo {year} {1980})}\BibitemShut {NoStop}%
\bibitem [{\citenamefont {Rao}(2005)}]{Rao2005}%
  \BibitemOpen
  \bibfield  {author} {\bibinfo {author} {\bibfnamefont {S.~S.}\ \bibnamefont
  {Rao}},\ }\href@noop {} {\bibinfo {title} {The finite element method in
  engineering, ed}} (\bibinfo {year} {2005})\BibitemShut {NoStop}%
\bibitem [{\citenamefont {Bathe}(1982)}]{Bathe1982}%
  \BibitemOpen
  \bibfield  {author} {\bibinfo {author} {\bibfnamefont {K.}~\bibnamefont
  {Bathe}},\ }\href@noop {} {\emph {\bibinfo {title} {Finite Element Procedures
  in Engineering Analysis}}}\ (\bibinfo  {publisher} {Prentice-Hall},\ \bibinfo
  {year} {1982})\BibitemShut {NoStop}%
\bibitem [{\citenamefont {Courant}\ and\ \citenamefont
  {Hilbert}(2008)}]{Courant2008MMP}%
  \BibitemOpen
  \bibfield  {author} {\bibinfo {author} {\bibfnamefont {R.}~\bibnamefont
  {Courant}}\ and\ \bibinfo {author} {\bibfnamefont {D.}~\bibnamefont
  {Hilbert}},\ }\href@noop {} {\emph {\bibinfo {title} {Methods of Mathematical
  Physics, Volume 1}}}\ (\bibinfo  {publisher} {Wiley},\ \bibinfo {year}
  {2008})\BibitemShut {NoStop}%
\bibitem [{\citenamefont {Nielsen}\ and\ \citenamefont
  {Chuang}(2010)}]{Nielsen2010QCQI}%
  \BibitemOpen
  \bibfield  {author} {\bibinfo {author} {\bibfnamefont {M.~A.}\ \bibnamefont
  {Nielsen}}\ and\ \bibinfo {author} {\bibfnamefont {I.~L.}\ \bibnamefont
  {Chuang}},\ }\href@noop {} {\emph {\bibinfo {title} {Quantum Computation and
  Quantum Information: 10th Anniversary Edition}}}\ (\bibinfo  {publisher}
  {Cambridge University Press},\ \bibinfo {address} {Cambridge},\ \bibinfo
  {year} {2010})\BibitemShut {NoStop}%
\bibitem [{\citenamefont {Zhang}\ \emph {et~al.}(2014)\citenamefont {Zhang},
  \citenamefont {Lu},\ and\ \citenamefont {Gao}}]{Zhang2014SCIS}%
  \BibitemOpen
  \bibfield  {author} {\bibinfo {author} {\bibfnamefont {Y.}~\bibnamefont
  {Zhang}}, \bibinfo {author} {\bibfnamefont {K.}~\bibnamefont {Lu}},\ and\
  \bibinfo {author} {\bibfnamefont {Y.}~\bibnamefont {Gao}},\ }\href
  {https://api.semanticscholar.org/CorpusID:255192859} {\bibfield  {journal}
  {\bibinfo  {journal} {Science China Information Sciences}\ }\textbf {\bibinfo
  {volume} {58}},\ \bibinfo {pages} {1 } (\bibinfo {year} {2014})}\BibitemShut
  {NoStop}%
\bibitem [{\citenamefont {Möttönen}\ \emph {et~al.}(2005)\citenamefont
  {Möttönen}, \citenamefont {Vartiainen}, \citenamefont {Bergholm},\ and\
  \citenamefont {Salomaa}}]{Mottonen2005JQIC}%
  \BibitemOpen
  \bibfield  {author} {\bibinfo {author} {\bibfnamefont {M.}~\bibnamefont
  {Möttönen}}, \bibinfo {author} {\bibfnamefont {J.~J.}\ \bibnamefont
  {Vartiainen}}, \bibinfo {author} {\bibfnamefont {V.}~\bibnamefont
  {Bergholm}},\ and\ \bibinfo {author} {\bibfnamefont {M.~M.}\ \bibnamefont
  {Salomaa}},\ }\href@noop {} {\bibfield  {journal} {\bibinfo  {journal} {J
  Quantum Info. Comput.}\ }\textbf {\bibinfo {volume} {5}},\ \bibinfo {pages}
  {467–473} (\bibinfo {year} {2005})}\BibitemShut {NoStop}%
\bibitem [{\citenamefont {Carrasquilla}\ \emph {et~al.}(2019)\citenamefont
  {Carrasquilla}, \citenamefont {Torlai}, \citenamefont {Melko},\ and\
  \citenamefont {Aolita}}]{Carrasquilla2019Nature}%
  \BibitemOpen
  \bibfield  {author} {\bibinfo {author} {\bibfnamefont {J.}~\bibnamefont
  {Carrasquilla}}, \bibinfo {author} {\bibfnamefont {G.}~\bibnamefont
  {Torlai}}, \bibinfo {author} {\bibfnamefont {R.~G.}\ \bibnamefont {Melko}},\
  and\ \bibinfo {author} {\bibfnamefont {L.}~\bibnamefont {Aolita}},\
  }\href@noop {} {\bibfield  {journal} {\bibinfo  {journal} {Nature Machine
  Intelligence}\ }\textbf {\bibinfo {volume} {1}},\ \bibinfo {pages} {155}
  (\bibinfo {year} {2019})}\BibitemShut {NoStop}%
\bibitem [{\citenamefont {Miller}\ \emph {et~al.}(2011)\citenamefont {Miller},
  \citenamefont {Wille},\ and\ \citenamefont {Sasanian}}]{Miller2011IEEE}%
  \BibitemOpen
  \bibfield  {author} {\bibinfo {author} {\bibfnamefont {D.~M.}\ \bibnamefont
  {Miller}}, \bibinfo {author} {\bibfnamefont {R.}~\bibnamefont {Wille}},\ and\
  \bibinfo {author} {\bibfnamefont {Z.}~\bibnamefont {Sasanian}},\ }in\ \href
  {https://doi.org/10.1109/ISMVL.2011.54} {\emph {\bibinfo {booktitle} {2011
  41st IEEE International Symposium on Multiple-Valued Logic}}}\ (\bibinfo
  {year} {2011})\ pp.\ \bibinfo {pages} {288--293}\BibitemShut {NoStop}%
\bibitem [{\citenamefont {Dekking}\ \emph {et~al.}(2006)\citenamefont
  {Dekking}, \citenamefont {Kraaikamp}, \citenamefont {Lopuha{\"a}},\ and\
  \citenamefont {Meester}}]{Dekking2006}%
  \BibitemOpen
  \bibfield  {author} {\bibinfo {author} {\bibfnamefont {F.}~\bibnamefont
  {Dekking}}, \bibinfo {author} {\bibfnamefont {C.}~\bibnamefont {Kraaikamp}},
  \bibinfo {author} {\bibfnamefont {H.}~\bibnamefont {Lopuha{\"a}}},\ and\
  \bibinfo {author} {\bibfnamefont {L.}~\bibnamefont {Meester}},\ }\href
  {https://books.google.com.sg/books?id=TEcmHJX67coC} {\emph {\bibinfo {title}
  {A Modern Introduction to Probability and Statistics: Understanding Why and
  How}}},\ Springer Texts in Statistics\ (\bibinfo  {publisher} {Springer
  London},\ \bibinfo {year} {2006})\BibitemShut {NoStop}%
\bibitem [{\citenamefont {Maslov}(2016)}]{Maslov2016PRA}%
  \BibitemOpen
  \bibfield  {author} {\bibinfo {author} {\bibfnamefont {D.}~\bibnamefont
  {Maslov}},\ }\href@noop {} {\bibfield  {journal} {\bibinfo  {journal}
  {Physical Review A}\ }\textbf {\bibinfo {volume} {93}},\ \bibinfo {pages}
  {022311} (\bibinfo {year} {2016})}\BibitemShut {NoStop}%
\bibitem [{\citenamefont {Broyden}(1970)}]{Broyden1970IMA}%
  \BibitemOpen
  \bibfield  {author} {\bibinfo {author} {\bibfnamefont {C.~G.}\ \bibnamefont
  {Broyden}},\ }\href@noop {} {\bibfield  {journal} {\bibinfo  {journal} {IMA
  Journal of Applied Mathematics}\ }\textbf {\bibinfo {volume} {6}},\ \bibinfo
  {pages} {76} (\bibinfo {year} {1970})}\BibitemShut {NoStop}%
\bibitem [{\citenamefont {Fletcher}(1970)}]{Fletcher1970TCJ}%
  \BibitemOpen
  \bibfield  {author} {\bibinfo {author} {\bibfnamefont {R.}~\bibnamefont
  {Fletcher}},\ }\href@noop {} {\bibfield  {journal} {\bibinfo  {journal} {The
  computer journal}\ }\textbf {\bibinfo {volume} {13}},\ \bibinfo {pages} {317}
  (\bibinfo {year} {1970})}\BibitemShut {NoStop}%
\bibitem [{\citenamefont {Goldfarb}(1970)}]{Goldfarb1970MOC}%
  \BibitemOpen
  \bibfield  {author} {\bibinfo {author} {\bibfnamefont {D.}~\bibnamefont
  {Goldfarb}},\ }\href@noop {} {\bibfield  {journal} {\bibinfo  {journal}
  {Mathematics of computation}\ }\textbf {\bibinfo {volume} {24}},\ \bibinfo
  {pages} {23} (\bibinfo {year} {1970})}\BibitemShut {NoStop}%
\bibitem [{\citenamefont {Shanno}(1970)}]{Shanno1970MOC}%
  \BibitemOpen
  \bibfield  {author} {\bibinfo {author} {\bibfnamefont {D.~F.}\ \bibnamefont
  {Shanno}},\ }\href@noop {} {\bibfield  {journal} {\bibinfo  {journal}
  {Mathematics of computation}\ }\textbf {\bibinfo {volume} {24}},\ \bibinfo
  {pages} {647} (\bibinfo {year} {1970})}\BibitemShut {NoStop}%
\bibitem [{\citenamefont {Ravi}\ \emph {et~al.}(2022)\citenamefont {Ravi},
  \citenamefont {Smith}, \citenamefont {Gokhale}, \citenamefont {Mari},
  \citenamefont {Earnest}, \citenamefont {Javadi-Abhari},\ and\ \citenamefont
  {Chong}}]{Ravi2022}%
  \BibitemOpen
  \bibfield  {author} {\bibinfo {author} {\bibfnamefont {G.~S.}\ \bibnamefont
  {Ravi}}, \bibinfo {author} {\bibfnamefont {K.~N.}\ \bibnamefont {Smith}},
  \bibinfo {author} {\bibfnamefont {P.}~\bibnamefont {Gokhale}}, \bibinfo
  {author} {\bibfnamefont {A.}~\bibnamefont {Mari}}, \bibinfo {author}
  {\bibfnamefont {N.}~\bibnamefont {Earnest}}, \bibinfo {author} {\bibfnamefont
  {A.}~\bibnamefont {Javadi-Abhari}},\ and\ \bibinfo {author} {\bibfnamefont
  {F.~T.}\ \bibnamefont {Chong}},\ }in\ \href@noop {} {\emph {\bibinfo
  {booktitle} {2022 IEEE International Symposium on High-Performance Computer
  Architecture (HPCA)}}}\ (\bibinfo {organization} {IEEE},\ \bibinfo {year}
  {2022})\ pp.\ \bibinfo {pages} {288--303}\BibitemShut {NoStop}%
\bibitem [{\citenamefont {Pellow-Jarman}\ \emph {et~al.}(2021)\citenamefont
  {Pellow-Jarman}, \citenamefont {Sinayskiy}, \citenamefont {Pillay},\ and\
  \citenamefont {Petruccione}}]{Pellow2021QIP}%
  \BibitemOpen
  \bibfield  {author} {\bibinfo {author} {\bibfnamefont {A.}~\bibnamefont
  {Pellow-Jarman}}, \bibinfo {author} {\bibfnamefont {I.}~\bibnamefont
  {Sinayskiy}}, \bibinfo {author} {\bibfnamefont {A.}~\bibnamefont {Pillay}},\
  and\ \bibinfo {author} {\bibfnamefont {F.}~\bibnamefont {Petruccione}},\
  }\href@noop {} {\bibfield  {journal} {\bibinfo  {journal} {Quantum
  Information Processing}\ }\textbf {\bibinfo {volume} {20}},\ \bibinfo {pages}
  {202} (\bibinfo {year} {2021})}\BibitemShut {NoStop}%
\bibitem [{\citenamefont {Wierichs}\ \emph {et~al.}(2020)\citenamefont
  {Wierichs}, \citenamefont {Gogolin},\ and\ \citenamefont
  {Kastoryano}}]{Wierichs2020PRR}%
  \BibitemOpen
  \bibfield  {author} {\bibinfo {author} {\bibfnamefont {D.}~\bibnamefont
  {Wierichs}}, \bibinfo {author} {\bibfnamefont {C.}~\bibnamefont {Gogolin}},\
  and\ \bibinfo {author} {\bibfnamefont {M.}~\bibnamefont {Kastoryano}},\
  }\href@noop {} {\bibfield  {journal} {\bibinfo  {journal} {Physical Review
  Research}\ }\textbf {\bibinfo {volume} {2}},\ \bibinfo {pages} {043246}
  (\bibinfo {year} {2020})}\BibitemShut {NoStop}%
\bibitem [{\citenamefont {Uvarov}\ \emph {et~al.}(2020)\citenamefont {Uvarov},
  \citenamefont {Biamonte},\ and\ \citenamefont {Yudin}}]{Uvarov2020PRB}%
  \BibitemOpen
  \bibfield  {author} {\bibinfo {author} {\bibfnamefont {A.}~\bibnamefont
  {Uvarov}}, \bibinfo {author} {\bibfnamefont {J.~D.}\ \bibnamefont
  {Biamonte}},\ and\ \bibinfo {author} {\bibfnamefont {D.}~\bibnamefont
  {Yudin}},\ }\href@noop {} {\bibfield  {journal} {\bibinfo  {journal}
  {Physical Review B}\ }\textbf {\bibinfo {volume} {102}},\ \bibinfo {pages}
  {075104} (\bibinfo {year} {2020})}\BibitemShut {NoStop}%
\bibitem [{\citenamefont {Sato}\ \emph {et~al.}(2023)\citenamefont {Sato},
  \citenamefont {Watanabe}, \citenamefont {Raymond}, \citenamefont {Kondo},
  \citenamefont {Wada}, \citenamefont {Endo}, \citenamefont {Sugawara},\ and\
  \citenamefont {Yamamoto}}]{Sato2023arXiv}%
  \BibitemOpen
  \bibfield  {author} {\bibinfo {author} {\bibfnamefont {Y.}~\bibnamefont
  {Sato}}, \bibinfo {author} {\bibfnamefont {H.~C.}\ \bibnamefont {Watanabe}},
  \bibinfo {author} {\bibfnamefont {R.}~\bibnamefont {Raymond}}, \bibinfo
  {author} {\bibfnamefont {R.}~\bibnamefont {Kondo}}, \bibinfo {author}
  {\bibfnamefont {K.}~\bibnamefont {Wada}}, \bibinfo {author} {\bibfnamefont
  {K.}~\bibnamefont {Endo}}, \bibinfo {author} {\bibfnamefont {M.}~\bibnamefont
  {Sugawara}},\ and\ \bibinfo {author} {\bibfnamefont {N.}~\bibnamefont
  {Yamamoto}},\ }\href@noop {} {\bibfield  {journal} {\bibinfo  {journal}
  {arXiv preprint arXiv:2302.12602}\ } (\bibinfo {year} {2023})}\BibitemShut
  {NoStop}%
\end{thebibliography}%
\bibliographystyle{apsrev4-2}
\end{document}